\documentclass[smallextended]{svjour3}     % onecolumn (second format)
\smartqed  % flush right qed marks, e.g. at end of proof
\usepackage{graphicx}
\usepackage{natbib}
\usepackage{multirow}
\usepackage{booktabs}
\usepackage[stable]{footmisc}
\newcommand{\otoprule}{\midrule[\heavyrulewidth]}
\begin{document}

\title{Pixel lensing}%\thanks{email:novati@sa.infn.it}}
\subtitle{Microlensing towards M31}

\titlerunning{Pixel lensing}        % if too long for running head

\author{S. Calchi Novati}

\authorrunning{S. Calchi Novati} % if too long for running head

\institute{S. Calchi Novati \at
              Dipartimento di Fisica ``E. R. Caianiello'', 
                Universit\`a di Salerno, 84084 Fisciano, Italy and
              Istituto Nazionale di Fisica Nucleare, Sezione di Napoli,
 Italy.\\
             \email{novati@sa.infn.it}           %  \\
}

\date{Received: 2 November 2009 / Accepted: 20 November 2009}

\maketitle

\begin{abstract}
Pixel lensing is gravitational microlensing of unresolved stars.
The main target explored up to now has been
the nearby galaxy of Andromeda, M31.
The scientific issues of interest 
are the search for dark matter in form of compact
halo objects, the study of the characteristics
of the luminous lens and source populations and
the possibility of detecting extra-solar (and extra-galactic) planets.
In the present work we intend to give an updated overview
of the observational status in this field.
\end{abstract}

\section{Introduction}
Following the original suggestion of Paczy\'{n}ski \cite{pacz86},
(stellar) gravitational microlensing
is by now an established and efficient tool of research.
The original motivation has been the search for dark matter in
Galactic halos in form of compact halo objects
(MACHOs). Meanwhile
microlensing probed to be a powerful
tool also for the analysis of the
characteristics of the (luminous) lens and source populations
and, more generally, of the Galactic structure. 
The first lines of sight to be explored
have been those towards the Magellanic Clouds
and the Galactic centre (for an updated account
see the review of Moniez \cite{moniez09}).
Along this second line of sight, the 
current main field of application is the search for
extra-solar planets (thoroughly discussed 
in the review of Dominik \cite{dominik09b}).

As for the search of compact halo objects,
the results obtained along
the line of sight towards the LMC are non-conclusive.
The MACHO collaboration claimed the detection of
a MACHO signal from objects of $\sim~0.4~\mathrm{M}_\odot$
that would constitute a (Milky Way) halo mass fraction
$f\sim 20\%$ \cite{macho00,bennett05}.
On the other hand, the EROS group
found no candidate events along this line of sight
and put a rather stringent limit \emph{upper}
limit, $f<0.1$, in the mass range of compact halo objects
preferred by the MACHO results \cite{eros07}.
More recently, the OGLE collaboration presented
two candidate events towards the LMC out of their OGLE-II
campaign, concluding that this is compatible
with the expected self-lensing signal \cite{ogle09}.

As soon as one wants to move beyond
these nearby targets a difficulty arises
in that the potential sources of microlensing
events are no longer resolved objects
and we enter the regime usually
referred to as \emph{pixel lensing}.
(At even larger distance we enter
the realm of quasar, or cosmological, gravitational microlensing,
which is the subject of the review
of Wambsganss \cite{wambsganss09}.)
Up to know the main field of application
of pixel lensing has been the nearby Andromeda galaxy, M31.

Pixel lensing, and in particular 
the line of sight towards the Andromeda galaxy,
is the subject of the present review.
Many reviews exist on the subject of gravitational microlensing
(e.g. Paczy\'{n}ski \cite{pacz96}, Roulet and Mollerach \cite{roulet97},
Wambsganss \cite{wambsganss06}). 
The theoretical aspects related more specifically
to pixel lensing have also been already
thoroughly discussed in a number of papers.
In the present work we intend to give an
updated overview of the current observational status in this field.
The outline is the following.
We start with a brief discussion of the
basic of pixel lensing, \S~\ref{sec:pl_theory}.
In \S~\ref{sec:m31} we discuss M31 pixel lensing:
we trace back the theoretical developments, 
\S~\ref{sec:m31_th}, and the observational campaigns
carried out along this direction, \S~\ref{sec:m31_obs}.
The modelling of M31 is discussed in \S~\ref{sec:m31_mod}
and the expected lensing signal in \S~\ref{sec:m31_exp}.
The main focus of the present review
is on the presentation of the observational results
obtained up to now along this line of sight, their interpretation
and the outlook for future
developments. This is the object of \S~\ref{sec:m31_res}.
In particular we present the candidate events in \S~\ref{sec:m31_evt}
and the results obtained on the MACHO content in \S~\ref{sec:m31_macho}.
In \S~\ref{sec:m31_eso} we discuss a further relevant
scientific application, the search for extra-solar planets
in M31 with pixel lensing.
Finally, in \S~\ref{sec:m31_beyond},
we discuss the application of pixel lensing
towards targets beyond the Local Group.

\section{Basic of Pixel lensing} \label{sec:pl_theory}

\emph{Pixel lensing is gravitational microlensing of unresolved stars} \cite{gould96}.
Looking for microlensing events, and
moving beyond the more nearby available targets
(the Galactic bulge and the Magellanic Clouds) the potential sources
are no longer resolved (though possibly blended) 
objects\footnote{Because of blending, often
the resolved microlensing sources are referred to as ``objects''
rather than ``stars''.}.
This establishes the difference with respect to
``classical'' gravitational lensing.
The key idea in this regime is therefore
to look for flux variations of the 
picture elements of the image (the pixels).
The appealing part of this approach is, 
besides the possibility to explore more distant targets,
the huge increase of potential sources
(all of them, more over, accessible to within
rather small field of views compared to local searches).
The related problem  is the impossibility, in most cases,
to access the source star flux. In turn, this 
enhances a series of problems in the analysis
peculiar to pixel lensing.

Quite generally, we may write the light curve flux for 
a microlensing event as
\begin{equation} \label{eq:light_curve}
\Phi\left( t,\{\theta\}\right) = 
\Phi^* \cdot \left(A\left(t,\{\theta\}\right)-1\right)+\Phi_\mathrm{B}\,.
\end{equation}
Here $\Phi^*$ is the flux of the unlensed source,
$A\left(t,\{\theta\}\right)$ is the
microlensing amplification, with $\{\theta\}$
being the amplification parameters,
and $\Phi_\mathrm{B}$ the \emph{background} flux level.
The microlensing amplification in the simplest
(and standard) situation of 
point source, point lens and 
uniform relative motion depends on three parameters: the 
impact parameter, $u_0$, the time of maximum
amplification, $t_0$, and the Einstein time,
$t_\mathrm{E}$ (the corresponding light curve
is usually referred to as \emph{Paczy\'{n}ski} light curve):
\begin{equation} \label{eq:at}
A\left(t,\{\theta\}\right) = \frac{u^2+2}{u \sqrt{u^2+4}}\,,
\end{equation}
where
\begin{equation} \label{eq:ut}
u\equiv u\left(t,\{\theta\}\right) 
= \sqrt{u_0^2+\left(\frac{t-t_0}{t_\mathrm{E}}\right)^2}\,.
\end{equation}
The relevant physical parameter is $t_\mathrm{E}$
which sets the fundamental timescale of the event.
It depends from the lens \emph{mass}, $M$,
the most relevant physical information one is 
interested into, 
and other usually non directly observable quantities
as the lens and source distances, $D_\mathrm{l}$ and $D_\mathrm{s}$
respectively, and the lens relative velocity
$v$ with respect to the line of sight, as $t_\mathrm{E} = R_\mathrm{E}/v$.
$R_\mathrm{E}$ is the Einstein radius,
and $\theta_\mathrm{E}=R_\mathrm{E}/D_\mathrm{l}$
the angular Einstein radius (on the lens plane)
\begin{equation} \label{eq:thetaE}
\theta_\mathrm{E} = \sqrt{\frac{4GM}{c^2} 
\frac{D_\mathrm{s}-D_\mathrm{l}}{D_\mathrm{l}\,D_\mathrm{s}}}\,.
\end{equation}
$\theta_\mathrm{E}$ is the fundamental (angular) length
scale which sets the cross-section of microlensing events.

In the ``classical'' lensing regime the quantity 
$\Phi(t)$ in Eq.~\ref{eq:light_curve}
is the measured flux from the lensed object.
With $A(t)\to 1$ for $|t|\gg t_0$,
the background level reads
$\Phi_\mathrm{B}=\Phi^*+\Phi_\mathrm{B}'$,
with $\Phi_\mathrm{B}'$ being the flux
of any unlensed source blended within the source star PSF.
In principle $\Phi_\mathrm{B}'$ is a well known
quantity so that the microlensing light curve
can be fitted to the four unknown parameters
$t_\mathrm{E},\,u_0,\,t_0$ 
and $\Phi^*$\footnote{From the observational point
of view the situation, however, is not as straightforward
as the actual blending fraction may not be as easy to be determined.
In fact, for microlensing analyses towards the Galactic bulge,
a robust agreement between theoretical expectations
and observational results has been reached 
only for a restricted
sample of selected bright, standard candle, sources for which
blending can be actually assumed to be negligible
\cite{popowski05,hamadache06,sumi06}.
Also for LMC analyses blending is a particularly
delicate issue to be dealt with \cite{macho00,eros07,ogle09}.}.

In the pixel-lensing regime, instead,
the quantity $\Phi(t)$ in Eq.~\ref{eq:light_curve}
is the \emph{pixel} flux. Within the same pixel
there is a large number of potential sources,
and an even much larger number of stars too faint
to give rise to any detectable lensing signal.
This can be looked at as the opposite case
of classical lensing, namely, a completely blended situation
with $\Phi_\mathrm{B}'\gg \Phi^*$. 
As an immediate consequence the photon \emph{noise} 
is going to be dominated by the underlying
background of the unresolved sources rather
than from the actual star being lensed
(and in particular of the amplification).
This is the characteristic signature
of pixel lensing analyses. 
A related issue is the
\emph{threshold} magnification  needed
to give rise to a detectable signal
that we are going to detail below.
As a direct consequence we may expect to be difficult
to measure from a light curve fit to the data
the unlensed flux and the event timescale, $t_\mathrm{E}$.
In fact, one usually finds a strong degeneracy
in the parameter space $t_\mathrm{E},\,u_0$ 
(or, equivalently, $\Phi^*,\,u_0$) \cite{wozniak_pacz87}. A possible way out
is that of reducing the number of free parameters from three to two 
$\Phi_*,\, u_0,\,t_\mathrm{E} \to \Delta\Phi,\,t_\mathrm{FWHM}$ \cite{wozniak_pacz87,gondolo99}.
Both the new parameters $\Delta \Phi$ and $t_\mathrm{FWHM}$
can always be easily measured on the light curve
as the flux difference at maximum amplification
and the full-width-half-maximum event duration, respectively.
The FWHM timescale, $t_\mathrm{FWHM}$,
is proportional to the Einstein timescale,
$t_\mathrm{FWHM} = t_\mathrm{E}\cdot f(u_0)$ \cite{gondolo99}.
For large values of the amplification, $u_0\ll 1$,
this relationship reduces to $t_\mathrm{FWHM} = \sqrt{12} u_0 t_\mathrm{E}$.

Gould \cite{gould96} made
the further distinction, within the pixel-lensing regime,
of a ``semiclassical'' regime,
where one can yet get to break
the parameter degeneracy and evaluate
the unlensed source  flux, and
a ``spike'' regime, where the background
level is so large that only events with
extremely large amplification, $u_0\ll 1$, are observable,
and where it is actually impossible to determine
the physical duration $t_\mathrm{E}$.

As remarked by Gould \cite{gould96},
M31 pixel lensing is at the limit
between the semiclassical and the spike regime.
It is not obvious to, but sometime
possible to measure out of the observed
light curves, with a reasonable
precision, the physical timescale $t_\mathrm{E}$.
To this purpose, an extremely good sampling
along the flux variation, and in particular
along their wings \cite{gould96,baltz00,dominik09a}, is essential.
Besides, a suitable sampling is in order
also to distinguish lensing signals from intrinsic variable objects.
As a note of terminology, therefore,
we may say that a ``pixel-lensing'' event,
besides the access to the knowledge of the source flux,
is a microlensing event for which 
the photon noise is dominated by that
of the background level.

In the classical regime a microlensing
event is considered to be enhanced whenever
the impact parameter, $u_0$, gets smaller than one.
For a specific experimental campaign
this value may be chosen to be 
somewhat smaller or larger, but in any case
it is a fixed quantity whatever, in particular,
the source flux value. In the pixel lensing
regime the situation is altogether different.
The threshold impact parameter, $u_\mathrm{T}$, is determined
given the, line of sight dependent, background noise
and depending on the underlying source luminosity function \cite{kerins01}.
In particular it turns out that, typically, 
$u_\mathrm{T} \sim {\cal O}(10^{-2} - 10^{-3})$.

To conclude, before moving to applications of pixel lensing
towards M31 and even more distant targets, we note
that pixel lensing can in fact be used also for \emph{nearby}
targets. In particular, microlensing analyses towards the LMC are usually
carried out looking at resolved objects, but 
one can expect microlensing events also from unresolved
LMC sources (in fact, as it turns out,
more events than from just resolved objects).
Such a programme has been carried out using
a part of the EROS-1 data towards the LMC bar \cite{melchior98,melchior99}.

\section{M31 pixel lensing} \label{sec:m31}

\subsection{Theoretical developments} \label{sec:m31_th}

The idea of looking for microlensing events
with sources (and lenses) belonging to M31 has
been first, independently, proposed by 
Crotts \cite{crotts92} and Baillon et~al. \cite{agape93}.
Crotts \cite{crotts92} first acknowledged the opportunity
given by the geometry of M31, the inclination of the M31
disk, to get a signature in the spatial distribution of microlensing events
due to M31 compact halo objects. Furthermore, 
he considered, as a possible detection
method, a first idea of what was going to become
the scheme of \emph{difference image photometry} \cite{tomaney96}.
On the other hand, the main focus of the analysis
presented by Baillon et~al. \cite{agape93}
has been on the acknowledgement  of the role
played by looking for microlensing events
due to unresolved stars. As discussed in the previous
Section, this introduces
additional difficulties in the analysis
but, on the other hand, translates
in an appealing substantial increase
into the number of potential sources.
Furthermore, they have presented
a first, detailed, Monte Carlo simulation framework
to establish the expected signal for a microlensing experiment.
In particular they have shown the relevant result
that we may expect the sources of most events to be 
bright stars (as a typical value, roughly $M_V<2$) with not 
so large amplification, and not, therefore,
faint sources with extremely large amplification.
Furthermore, following this first analysis, an original scheme 
of light curve analysis, the \emph{superpixel photometry}
\cite{agape97,melchior99}\footnote{Difference image analysis
and superpixel photometry are still the two 
currently used  photometry analysis schemes for pixel lensing observations.
In fact these two methods
are not mutually exclusive, in that a possible,
and perhaps suitable strategy, would be
to use the superpixel photometry approach
for the identification of flux variations, and then
make use of the more refined difference image photometry
only on a subset of selected light curves.}
was then developed. 
Shortly after, Jetzer \cite{jetzer94} presented
a more detailed theoretical study
on gravitational microlensing towards
M31 with emphasis on the relevant
microlensing quantities, the \emph{optical depth}
and the \emph{microlensing rate}.

A further advantage of the line of sight
towards M31 for MACHO searches, whose relevance can not be
stressed enough, is the possibility, looking at it from outside, 
to fully map the M31 own dark matter halo.
Such an analysis is not possible
for the Galactic halo.

These first theoretical analyses have been
thoroughly developed. Colley \cite{colley95}
have formalized the problem of detecting events
through the ``threshold'' approach.
Han \cite{han96} analysed the problem
of observations towards the M31 bulge
evaluating the optical depth, the 
timescale distribution and the expected
event rate for a specific observational set up.
Furthermore, he provided an important
result on the extinction by the dust in the M31 disk.
Gondolo \cite{gondolo99} considered
the problem of evaluating the optical depth
in the pixel lensing regime, namely,
by making use of the FWHM timescale, $t_\mathrm{FWHM}$,
rather than the Einstein timescale, $t_\mathrm{E}$.
Further analyses concerning the optical depth
and the event rate for M31 pixel lensing
have been carried out by Gyuk and Crotts \cite{gyuk_crotts00}
and by Baltz and coauthors
\cite{baltz00,baltz03,baltz05}.

A crucial aspect of any microlensing campaign
is the estimate of the expected signal.
Kerins et~al. \cite{kerins01} outlined
a detailed scheme for a pixel lensing simulation.
Riffeser et~al. \cite{arno06}
presented a thorough analysis
of the theory and application of microlensing towards M31.
The scheme of the Monte Carlo simulation
first sketched in \cite{agape93,agape97} has been further
discussed by Calchi~Novati et~al. \cite{novati05,novati09}.

\subsection{Observational campaigns} \label{sec:m31_obs}

Following, and parallel, to these theoretical
developments, several observational pixel-lensing campaigns
have been undertaken towards M31.
The theoretical analyses of Crotts lead to 
the Vatican Advanced Technology Telescope[VATT]/Columbia campaign \cite{crotts96,uglesich04},
using data from the 1.3m Vatican telescope and the
MDM 1.3m telescope and to its successor,
the MEGA (Microlensing Exploration of the Galaxy and Andromeda)
collaboration who used the 2.5m INT (Isaac Newton Telescope)
\cite{mega04,mega06}. The work of Baillon et~al. \cite{agape93}
led to the formation of the 
AGAPE (Andromeda Galaxy and Amplified Pixels Experiment) 
collaboration\footnote{A brief history of
the beginning of the AGAPE collaboration is given in the Appendix.}
who used the 2m TBL (Telescope Bernard Lyot) \cite{agape97,agape99}.
As a part of this same project, within the SLOTT 
(Systematic Lensing Observations at Toppo Telescope)-AGAPE collaboration,
Calchi~Novati et~al. \cite{novati02,novati03}, 
thanks to a collaboration with A.~Gould 
and sharing data with the VATT/Columbia project,
analysed data collected at the 1.3m MDM telescope. 
Still from a collaboration with
the AGAPE group, the Nainital group
carried out a campaign with the 104cm Sampurnanad telescope \cite{joshi05}.
Eventually, from the AGAPE experiment,
the POINT-AGAPE (Pixel-lensing Observations with the Isaac Newton Telescope)
collaboration was begun sharing data collected at the 2.5m INT telescope
with the MEGA collaboration \cite{point01,paulin03,belokurov05,novati05}.
Finally, the WeCAPP (Wendelstein Calar Alto Pixellensing Project)
group carried out a several-years campaign
using both the 1.3m Calar Alto telescope and the 0.80m Wendelstein telescopes
\cite{wecapp01,wecapp03}. In Table~\ref{tab:m31_obs} we resume
the present status. Most observational campaigns have been observing
around the M31 bulge region, although usually avoiding
the very M31 centre. In Fig.~\ref{fig:m31_evt} we draw
the contours of the field observed by the POINT-AGAPE/MEGA collaborations,
up to now the pixel lensing observational campaign that covered 
the largest field of view.

\begin{table}
\caption{Completed and ongoing microlensing campaigns towards M31.
Fifth column: number of microlensing candidate events,
in bracket those that are no longer considered as such
(see text for details) and in particular not reported in Table \ref{tab:evt_obs}
and in Fig.~\ref{fig:m31_evt}.}
\label{tab:m31_obs} 
\begin{tabular}{cccccc}
\toprule
collaboration & years & telescope & f.o.v. & \# events & ref\\
\otoprule
\multirow{2}*{VATT/Columbia}
       & 1997 & 1.8m VATT &  $2\times (11.3'\times 11.3')$
& \multirow{2}*{(3)+3} & \multirow{2}*{\cite{crotts96,uglesich04}}\\
\cmidrule(l){2-4}
       &1997-1999 & 1.3m MDM & $2\times (17'\times 17')$ \\ 
\midrule
MEGA & 1999-2002 & 2.5m INT & $2\times(33'\times 33')$ & (4)+14 & \cite{mega04,mega06}\\
\midrule
AGAPE  & 1994-1996 & 2.0m TBL & $6\times (4.5'\times 4.5')$ & 1 & \cite{agape97,agape99} \\
\midrule
SLOTT-AGAPE & 1997-1999 & 1.3m MDM & $2\times (17'\times 17')$ & (5)+3 & \cite{novati02,novati03}\\ 
\midrule
POINT-AGAPE & 1999-2001 & 2.5m INT & $2\times(33'\times 33')$ & 7 & \cite{point01,paulin02,paulin03,belokurov05,novati05}\\
\midrule
\multirow{2}*{WeCAPP} 
       & \multirow{2}*{1997-2008} 
& 1.23m CA & $17.2'\times 17.2'$ & \multirow{2}*{2} & 
\multirow{2}*{\cite{wecapp01,wecapp03}}\\
\cmidrule(l){3-4}
       && 0.80m We & $8.3'\times 8.3'$ \\ 
\midrule
Nainital & 1998-2002 & 1.04m Sa& $13'\times 13'$ & 1 & \cite{joshi05}\\
\midrule
ANGSTROM & 2004- & 2m LT \& FTN & $4.6'\times 4.6'$ & - &\cite{kerins06}\\
\midrule
PLAN & 2006- & 1.5m OAB & $2\times (13'\times 12.6')$ & 2 & \cite{novati07,novati09}\\
\bottomrule
\end{tabular}
\end{table}

Ongoing microlensing campaigns are trying to 
overcome some of the problems of the first campaigns.
E.~Kerins et~al. \cite{kerins06} started the ANGSTROM 
(Andromeda Galaxy Stellar Robotic Microlensing) collaboration 
to probe stellar lensing in the inner bulge region of M31
down to low mass stars with the specific aim
to constrain the 3d structure of the M31 bulge.
By making primarily use of a network of 2m class robotic telescopes
with a small field of view, $4.6'\times 4.6'$, centered
right in the M31 centre (the Liverpool Telescope, LT, 
and the Faulkes Telescope North, FTN), plus 3 additional
telescopes (the 1.8m Bohyunsan Observatory in Korea,
the 2.4m Hiltner MDM at Kitt Peak, US, and the 1.5m
Maidanak Observatory in Uzbekistan),
the ANGSTROM collaboration
has also undertaken the first ambitious and challenging project
of real time microlensing discovery outside the Galaxy: the 
APAS, the Angstrom Project Alert System \cite{darnley07}.
In particular they also considered the possibility
to use this high cadence survey to flag and follow
up binary systems in M31 \cite{kim07}.
The ANGSTROM project has begun taking data in 2004.
The PLAN (Pixel Lensing Andromeda) collaboration 
undergone a new observational campaigns 
making use of the 1.5m Loiano Telescope at OAB (Astronomical
Observatory of Bologna, Italy)
\cite{novati07}. With a CCD field of view of $13'\times 12.6'$
they have been monitoring two fields around the inner M31 region,
with an aggressive observational strategy
of consecutive and full nights. Given
the short duration of the expected events,
a good sampling is essential first
to safely distinguish microlensing events
from intrinsic variables and then
to robustly characterize the detected events.
The PLAN collaboration started observing with
a pilot season campaign in 2006, and is still
currently carryng on his observational effort.
In particular they have reported the detection
of 2 microlensing candidates from their 2007 season \cite{novati09}.
In 2008 they begun using also the 1.5m TT1 telescope
(Astronomical Observatory of Capodimonte, Napoli, Italy).

Both these ongoing observational campaigns suffer,
in the perspective of the search for MACHOs,
from an intrinsic limitation in that
their observed fields of view are rather small.
As we will detail later, the spatial distribution
of the lensing events is an important issue, 
so that the possibility to monitor not only
the M31 central region is crucial 
for a correct understanding of the lensing signal.
The PAnandromeda project\footnote{S. Seitz, talk given at the 13th Microlensing
Workshop, 2009, Paris.}  plans to make use
of the 1.8m PS1 telescope\footnote{http://pan-starrs.ifa.hawaii.edu/public/} with a huge
field of view of 6.4 sqdeg (with an almost
complete $64\times 64$ array of CCD devices,
each about $600\times 600$ pixels and 
using the ``orthogonal transfer'' technique for the read out). 
This will cover in a single shot all of the M31 field.
M31 is expected to be monitored with a cadence
of nightly exposures (such a tight sampling
is extremely important, indeed essential, to cope with the
expected flux variations lasting a few days only) of 12m and 6m in $r'$ and $i'$
band, respectively, for about 10 weeks per season
(this overall exposure time per night, however,
given the mirror size, may not allow
to go fainter than previous campaigns,
in particular the INT one which, with a 2.5m telescope,
observed each field, usually, about 20m per night in $r'$ band,
though with a much more irregular sampling).
Such an ambitious project is expected 
to bring extremely exciting results on M31 pixel lensing.
The observational campaign has been started in 2009.

\subsection{Modelling of M31} \label{sec:m31_mod}

A correct modelling of M31 is an essential ingredient
for any analysis aiming at describing the expected
lensing signal along this line of sight.
The starting point is an accurate model for 
the luminous components responsible for the
expected self-lensing signal. This is particularly
delicate also because, opposite to the LMC case,
this signal, at least in the central M31 region,
is in fact comparable to the would be MACHO signal.

M31 lies at a distance estimated at $\sim 785~\mathrm{kpc}$ \cite{mcconnachie05}
(see also \cite{tammann08} who proposed a somewhat larger value),
with a sharp inclination angle of $77^\circ$ \cite{walterbos87}.
The morphology of M31 is similar to that of the Milky Way,
with a central bulge and a disc. 
The single one fundamental physical parameter linked
to the expected lensing signal is the overall \emph{stellar} mass
of these luminous components. As for the bulge mass
the estimate of Kent \cite{kent89} of $4.0\times 10^{10} \mathrm{M}_\odot$
(for which a distance to M31 of ``only'' $690~\mathrm{kpc}$ was used)
has become a sort of ``standard'' reference value
for most microlensing analyses \cite{gyuk_crotts00,kerins01,baltz03,arno06,novati09}.
Together with a disc mass of $3.1\times 10^{10} \mathrm{M}_\odot$ 
\cite{kerins01,arno06,novati05,novati09} this gives
an overall M31 stellar mass of $\sim 7 \times 10^{10} \mathrm{M}_\odot$.
For their POINT-AGAPE analysis, however, Calchi~Novati et~al. \cite{novati05}
used, as a fiducial value, a lighter bulge of $1.5\times 10^{10} \mathrm{M}_\odot$,
and the Kent \cite{kent89} value as a test model only.
Finally, de~Jong et~al. \cite{mega06}, for their MEGA analysis, 
used, for their fiducial model, 4.4 and 5.5 (in units of $10^{10} \mathrm{M}_\odot$) 
for the bulge and the disc mass, respectively.
On the other hand, most non-microlensing analyses usually
get to a somewhat larger overall stellar mass, $\sim 10 \times 10^{10} \mathrm{M}_\odot$,
but with a smaller bulge mass. In particular  bulge and disc mass values
are reported to be (in units of $10^{10} \mathrm{M}_\odot$) 2.4 and 7.1 \cite{chemin09},
3.2 and 7.2 \cite{geehan06}, 2.5 and 7.0 \cite{widrow03}, 3.4 and 5.6 \cite{tamm07b},
1.9 and 7.0 \cite{klypin02}. Clearly there is still not a consensus
on this issue. One leading reason is the uncertainty linked
to the  $M/L$ ratio values and  to the (related) issue
of the internal M31 extinction (see also \cite{montalto09}).
It is beyond our scope to enter a detailed
discussion about this matter for which we refer to the previously cited works.
We only note that, in particular for the bulge mass,
the rather large value in microlensing analyses is usually assumed 
to be on the ``safe'' side in order not to underestimate the lensing bulge contribution.
On the other hand, in microlensing analyses one should be careful
to actually take into account the stellar mass contribution only,
and this may be at odd with determinations based on dynamical grounds.
Clearly, a correct estimate of the bulge and disc mass
is essential in order to get to reliable estimates
of the expected self-lensing signal.

For the overall bulge and disc structure the models are based upon
the available surface brightness profiles. Besides the work
of Walterbos and Kennicutt \cite{walterbos87}, we recall
also the analyses of Kent \cite{kent83,kent86,kent87} and in particular
the bulge/disc decomposition discussed in \cite{kent89} that have been 
taken as a basis, in particular, for many microlensing analyses
\cite{kerins01,novati05,arno06,novati09}. Noteworthy, to discuss
their microlensing results, de~Jong et~al. \cite{mega06} 
made use of a fully self-consistent M31 model following
the analysis of Widrow and Dubinksi \cite{widrow05}.
More detailed morphological analyses of the inner M31 bulge
suggest also the presence of a bar-like structure \cite{athanassoula06,beaton07}.
A specific analysis on this issue from a microlensing
perspective is given in Kerins et~al. \cite{kerins06}.

As for the bulge structure and population we finally mention
the recent work of Saglia et~al. \cite{saglia09} who present and 
discuss new optical long-slit data out to $5'$ from the M31 centre
with the purpose to constrain the stellar and gas kinematics. 
In particular they point out that previous estimates
of the velocity dispersion were severely underestimated
and this lead them to revise upward, with respect
for instance to the analyses in \cite{widrow03,arno06,chemin09}, the stellar mass of the bulge.
Their analysis also suggests the possibility
of an intrinsic triaxiality of the bulge and/or the presence of a bar. 

A further, somewhat elusive, stellar component is the stellar halo of M31
\cite{font08,tanaka09}, known in particular to be rich 
in substructures \cite{ferguson02,richardson08}.
The microlensing signal from this component may be expected to be enhanced
because of the  increase in the lens-source distance,
however, because of the small density
and overall total mass, one can expect that this should not be
large with respect to microlensing events by lenses
in the bulge or the disc.  Anyway, a 
specific analysis on this issue still awaits to be carried out.

In order to specify a microlensing event one
also needs the mass function for the lenses
(given the overall mass of a lens component, this is
related to the number of available lenses)
and the luminosity function for the sources
(given the surface brightness,
this is related to the number of available sources along a given line of sight).
We still lack definitive results on both these issues, 
even if the content of the M31 stellar bulge and disc
have been already the object of several analyses \cite{rich95,stephens03,sarajedini05,olsen06,yin09}
(for the mass function we recall in particular the work of \cite{ballero07}). 
Therefore, most times Milky Way results (for the mass function)
and synthetic luminosity functions are used (for microlensing-based analyses 
we refer for instance to the discussions in Riffeser et~al. \cite{arno06} and in 
Kerins et~al. \cite{kerins06}). 
A relevant aspect to stress is that, whereas
for the mass function one is mostly interested in the low mass tail, 
where most lenses are,
for the luminosity function  one is mainly interested in the opposite tail, 
the bright end, as one expect only giant
stars as possible sources.
Finally, one needs models for the velocity distributions 
(Ref.~\cite{saglia09} and references therein)
and the M31 transverse velocity \cite{vdm08}.

In an early work, Braun \cite{braun91}, with a study 
of the neutral gas in M31, and in particular
of the rotation curve out to $28~\mathrm{kpc}$, concluded
on the lack of indications for a massive dark halo.
This view is by now altogether changed. 
For instance, in a recent analysis, Chemin et~al. \cite{chemin09},
by studying the M31 rotation curve out to $38~\mathrm{kpc}$,
conclude that the dark matter component is almost 4 times more massive
than the baryonic mass. The authors also consider
different shape for the dark matter halo, the Navarro-Frenk\& White 
\cite{nfw96,nfw97} model,
the Einasto model \cite{merritt06}, and the (pseudo-)isothermal sphere.
Noteworthy, these models all fail to exactly reproduce the observed rotation curve,
and neither of them is found to be preferred.

\subsection{The expected signal} \label{sec:m31_exp}

A first characterisation of the expected
pixel lensing signal may come from
an analysis of the microlensing quantities,
the optical depth and the microlensing rate.
However, to properly address this issue
a full simulation of a given experiment is needed.
This holds because of the interplay
among the background noise level,
the threshold impact parameter and the
luminosity function (more specifically,
the exact fraction of unresolved
stars that must be counted as possible sources) 
\cite{kerins01,arno06,kerins06,novati05,novati09}.
In particular, a fundamental issue to be addressed is the 
correct understanding of the nature of the lenses.
Either belonging to some known luminous population,
for self-lensing events, or to the
dark matter halo, for MACHO lensing. 
This is never trivial for microlensing events, 
as lens mass, lens and source distances and relative
velocities are not directly measured. In the  pixel lensing regime
this is further complicated because of the additional
degeneracy due to the ignorance of the source flux.
Self lensing constitutes a background one must
get rid of for the study of the MACHO lensing signal.
Nonetheless, the  self-lensing signal is relevant in itself.
It allows one to assess the capability
of one's pipeline to detect microlensing events,
whenever expected, and, if possible, to
``normalize'' the expected self-lensing versus MACHO lensing signal.
Furthermore, self lensing can also be used
to study the characteristics of the luminous lens populations
(for Galactic bulge observations we recall, for instance, the analysis
of the bulge mass function in \cite{novati08}). 

At a first level of analysis three main statistics
are available to characterize the lensing signal:
the number of observed events, the duration and
the distance from the M31 centre distributions.
In principle, as for the spatial distribution,
one might want to study the expected asymmetry
for lensing events due to compact objects belonging
to the M31 halo (Milky Way
compact halo objects are expected to
contribute, for given MACHO mass and halo fraction,
for about one third of the overall MACHO signal \cite{arno06}).
To do so, however, a rather large number
of events is necessary, and indeed the MEGA collaboration
studied this statistics \cite{mega06}. 
Further care in this analysis is however suggested by the results by An et~al. \cite{an04}
who have shown that differential extinction
in M31 might induce a similar asymmetric signal
on the spatial distribution of self-lensing events.
The distance from the M31 centre can be taken, in any case,
as a useful zeroth order approximation \cite{novati05,novati07}
as self-lensing events are expected to be more clustered
around the M31 centre.
A fourth available statistics is the flux deviation at maximum.
However, this is only useful as a consistency check for the analysis,
but it contains no information about the nature of the lenses.

\begin{figure}
\includegraphics[width=120mm]{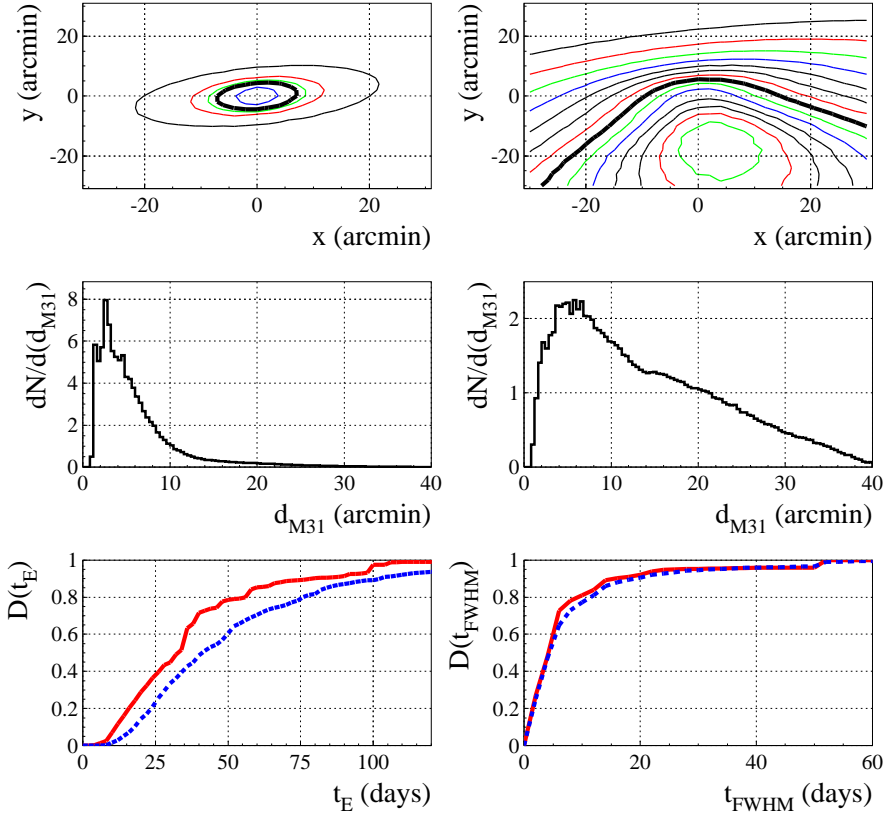}
\caption{The expected signal. Top panels: Microlensing rate per year per source star
with $u_\mathrm{T}=\mathrm{cost}=1$, for bulge sources and bulge (left)
and M31 MACHO of $0.5~\mathrm{M}_\odot$ lenses. The $x-y$ axes are 
given in an intrinsic M31 coordinate system. The M31 models
are as in \cite{arno06}. The contour levels are $0.1,0.4,0.7,1.0,2.5$
(left panel) and from 0.3 to 1.6 with $\Delta=0.1$ with
the thicker line marking the $1.0$ level
(always in units of $10^{-5}~\mathrm{events}~\mathrm{yr}^{-1}$).
Middle panels: Distance from the M31 centre distribution
for self-lensing (left) and MACHO lensing. These distributions,
with arbitrary normalization,
are taken as an output of the Monte Carlo simulation
of the POINT-AGAPE group presented in \cite{novati05}.
Bottom panels: Cumulative distributions
for $t_\mathrm{E}$ (left) and $t_\mathrm{FWHM}$
for bulge-bulge (solid lines) and M31 MACHO of $0.5~\mathrm{M}_\odot$.
These distributions are taken as an output
of the Monte Carlo simulation presented in \cite{novati09}.
}
\label{fig:exp}    
\end{figure}

The \emph{number} of the expected events, in the central M31 region,
from self lensing and MACHO lensing are about of the same order.
The exact number depends on the M31 model,
the field of view and the observational  set up
(besides the MACHO mass and halo fraction).
We can take as an example the results reported in the
analysis of Riffeser et~al. \cite{arno06} in their Table~2,
for the expected signal in a field of view of about $17'\times 17'$
around the M31 centre. To be more specific, 
let consider compact halo objects of $0.5~\mathrm{M}_\odot$
and an halo mass fraction $f=20\%$,
about the preferred values from the MACHO LMC analysis \cite{macho00}.
It then turns out that the number of expected self-lensing events
is indeed larger, by about a factor of 1.5, 
than that of MACHO lensing ones\footnote{This outcome 
marks a relevant difference with respect to the LMC analyses where, even
when considering the central bar region only,
the number of expected MACHO lensing events
outcomes that of self lensing (even if the exact
contribution of LMC self lensing is still a debated issue
\cite{gyuk00,jetzer02,mancini04,novati06,novati09b}).}.

This result can also be deduced by the top panels
of Fig.~\ref{fig:exp} where we report
the microlensing rate 
for bulge sources and bulge and M31 $0.5~\mathrm{M}_\odot$
compact halo objects lenses (using the same models of \cite{arno06}).
Here the rate is evaluated per source star and
for a fixed threshold impact parameter, $u_\mathrm{T}=1$,
so that it can not be directly compared to the observations.
However it clearly shows, first, the already stated asymmetry
expected for M31 halo lensing events. Second, that self lensing, at least
in the central region, is expected to give rise
to a similar signal, as for the number of events, than MACHO lensing
(note in particular the position of the line of equal rate at 
$10^{-5}~\mathrm{events}~\mathrm{yr}^{-1}$).

To evaluate the actual number of expected events
this expression of the rate must be multiplied,
roughly the same overall factor  for the different lens populations, 
for the number of available sources using the appropriate value
for the threshold impact parameter, $u_\mathrm{T}$. 
Depending on the luminosity function, the number of available
sources follows the M31 surface brightness
profile and peaks at the M31 centre.
This explains the increase towards the M31 centre of the expected number of events
for all the lens populations.
On the other hand,  the threshold value for the impact  parameter, $u_\mathrm{T}$, 
is a function of the line of sight (\S ~\ref{sec:pl_theory}). 
Corresponding to the increase in the noise level 
following the M31 surface brightness,
the threshold impact parameter decreases.
As a final outcome, the expected number of lensing events, 
for both MACHO lensing and self lensing,
moving towards the M31 centre first increases and then decreases
with a turn off point determined by the specific observational set up.
In Fig.~\ref{fig:exp}, middle panels, we show
the results of the Monte Carlo simulation for the POINT-AGAPE experiment,
\cite{novati05}, for the distribution
of the expected distance from the M31 centre.
Finally, moving towards the M31 centre, one
must also face the problem of the increasing
crowding of the field which further decreases
the expected signal. To properly evaluate
the extent of this effect is one
of the main purposes of the detection 
efficiency analyses.

This result motivates the need of looking for microlensing events
also in the outer regions of M31, where the expected
self-lensing signal may be considered negligible.
On the other hand, this is also where the expected 
overall rate becomes quite small, 
not a secondary issue as a main problem for the interpretation
of the current experiments is the overall low rate of observed events.
The analysis of the inner M31 region remains however
essential, as the observation (or not) of 
the expected self-lensing signal
may be used as a ruler for a given experiment.

The further relevant statistics
we have from the observations is the event duration.
As discussed in \S~\ref{sec:pl_theory}, the directly
accessible quantity is the full-width-half-maximum duration,
$t_\mathrm{FWHM} = t_\mathrm{E} \cdot w(u_0)$.
Once again, the threshold impact parameter,
and its dependence on the background noise level, enters into play.
In particular, at least for MACHO mass in the
same range of stellar masses, the differences that exist
in the physical duration distributions
for MACHO lensing and self-lensing
are almost completely washed out.
It also turns out that, whatever the lens population,
most of the events are  expected to last a few days only.
In Fig.~\ref{fig:exp}, bottom panels, we show the cumulative
distribution for the Einstein time and the full-width-halo-maximum
timescales for the bulge sources and bulge and $0.5~\mathrm{M}_\odot$
M31 MACHO lenses configurations
from the Monte Carlo simulation of the Loiano experiment
discussed by the PLAN collaboration \cite{novati09}.

At a more refined level of analysis,
a possible expected deviation
from the standard Paczy\'{n}ski light curve
that can be exploited for a deeper understanding
of the lens nature is
the finite size of the sources \cite{witt_mao_94}.
Indeed, this effect can become relevant as most sources
are expected to be bright giant stars, and in particular
it has been shown to be suitable to distinguish
MACHO lensing from self-lensing events \cite{arno08}.
Furthermore, deviations from the point-like source light curve shape
may also be used to constrain the lens proper motion \cite{gould94}.
In turn, this is relevant as it may allow one
to distinguish M31 lenses from Milky Way ones \cite{hangould96b,point01}.

\subsection{Observational results} \label{sec:m31_res}
\subsubsection{The microlensing candidate events}  \label{sec:m31_evt}

\begin{table}
\caption{Candidate microlensing events reported towards M31,
as shown in Fig.~\ref{fig:m31_evt}.
Seventh column: `$*$' indicates that the colour is $V-R$,
`$**$' indicates that the colour is $B-R$.
} \label{tab:evt_obs} 
\begin{tabular}{rllrrclll}
\hline
id & RA & DEC & $d_\mathrm{M31}$ & $t_\mathrm{FWHM}$ & ${\Delta}R_\mathrm{MAX}$ &
$R-I$ & name  & REF\\
& (deg) & (deg) & (arcmin) & (days) &&&&\\
\noalign{\smallskip}\hline\noalign{\smallskip}
 1&10.672917&41.277528&  0.72&  5.3 &  17.9 & 0.8(**) & AGAPE-Z1 & \cite{agape99}\\
 2&10.713333&41.398972&  7.89&  1.8 &  20.8 & 1.2(*)  & PA-N1/MEGA-ML16 &\cite{point01,mega06}\\
 3&11.087083&41.479111& 22.07& 22.0 &  19.1 & 1.0(*)  & PA-N2/MEGA-ML7 &\cite{paulin03,mega04}\\
 4&10.626250&41.216833&  4.10&  2.3 &  18.8 & 0.6  & PA-S3/GL1 &\cite{paulin03,wecapp03,belokurov05}\\
 5&10.625000&40.896139& 22.55&  2.0 &  20.7 & 0.0  & PA-S4/MEGA-ML11 &\cite{paulin02,mega04,belokurov05}\\
 6&10.552083&41.358333&  8.01& 16.0 &  21.0 & 2.2  & C3 &\cite{novati03}\\
 7&10.606667&41.440833& 10.87& 13.0 &  21.3 & 1.1  & C4 &\cite{novati03}\\
 8&10.470000&41.288333&  9.74& 14.0 &  21.8 & 0.5  & C5 &\cite{novati03}\\
 9&10.636667&41.332361&  4.36&  5.4 &   -   & 1.1  & GL2 &\cite{wecapp03}\\
10&10.845417&41.091667& 12.90& 26.5 &  22.2 & -    & 97-1267 &\cite{uglesich04}\\
11&10.762083&41.120694&  9.58& 17.3 &  20.3 & -    & 97-3230 &\cite{uglesich04}\\
12&10.988750&41.198972& 14.36&  2.2 &  21.8 & -    & 99-3688 &\cite{uglesich04}\\
13&10.793750&41.296611&  5.19&  5.4 &  21.8 & 0.6  & MEGA-ML1 &\cite{mega04}\\
14&10.799583&41.295444&  5.42&  4.2 &  21.5 & 0.3  & MEGA-ML2 &\cite{mega04}\\
15&10.815833&41.347833&  7.56&  2.3 &  21.6 & 0.4  & MEGA-ML3 &\cite{mega04}\\
16&10.852083&41.630667& 22.95& 27.5 &  22.3 & 0.6  & MEGA-ML8 &\cite{mega04}\\
17&11.195000&41.685194& 33.90&  2.3 &  22.0 & 0.2  & MEGA-ML9 &\cite{mega04}\\
18&10.978750&41.175917& 14.41& 44.7 &  22.2 & 1.1  & MEGA-ML10 &\cite{mega04}\\
19&10.760417&40.752556& 31.19& 26.8 &  23.3 & 0.8  & MEGA-ML13 &\cite{mega04}\\
20&10.927083&40.709417& 35.34& 25.4 &  22.5 & 0.4  & MEGA-ML14 &\cite{mega04}\\
21&10.509583&40.909722& 22.98&  3.4 &  19.5 & -0.1 & (PA-S16) & \cite{belokurov05}\\
22&10.544583&41.329278&  7.27&  1.8 &  20.8 & 0.5  & PA-N6 &\cite{novati05}\\
23&10.677500&41.211889&  3.46&  4.1 &  20.8 & 0.8(*)  & PA-S7 &\cite{novati05}\\
24&10.888750&41.128889& 12.49& 59.0 &  20.1 & 1.3  & NMS-E1 & \cite{joshi05}\\
25&10.788750&41.348167&  6.67& 16.1 &  21.6 & 0.5    & MEGA-ML15 &\cite{mega06}\\
26&10.481667&40.938889& 21.85& 10.1 &  22.2 & 0.4  & MEGA-ML17 &\cite{mega06}\\
27&10.822083&41.037139& 15.25& 33.4 &  22.7 & 0.5  & MEGA-ML18 &\cite{mega06}\\
28&10.737500&41.380556&  7.09&  7.1 &  21.1 & 1.0  & OAB-N1 &\cite{novati09}\\
29&10.708333&41.311111&  2.73&  2.6 &  19.1 & 1.1  & OAB-N2 &\cite{novati09}\\
\noalign{\smallskip}\hline
\end{tabular}
\end{table}

\begin{figure}
\includegraphics[width=120mm]{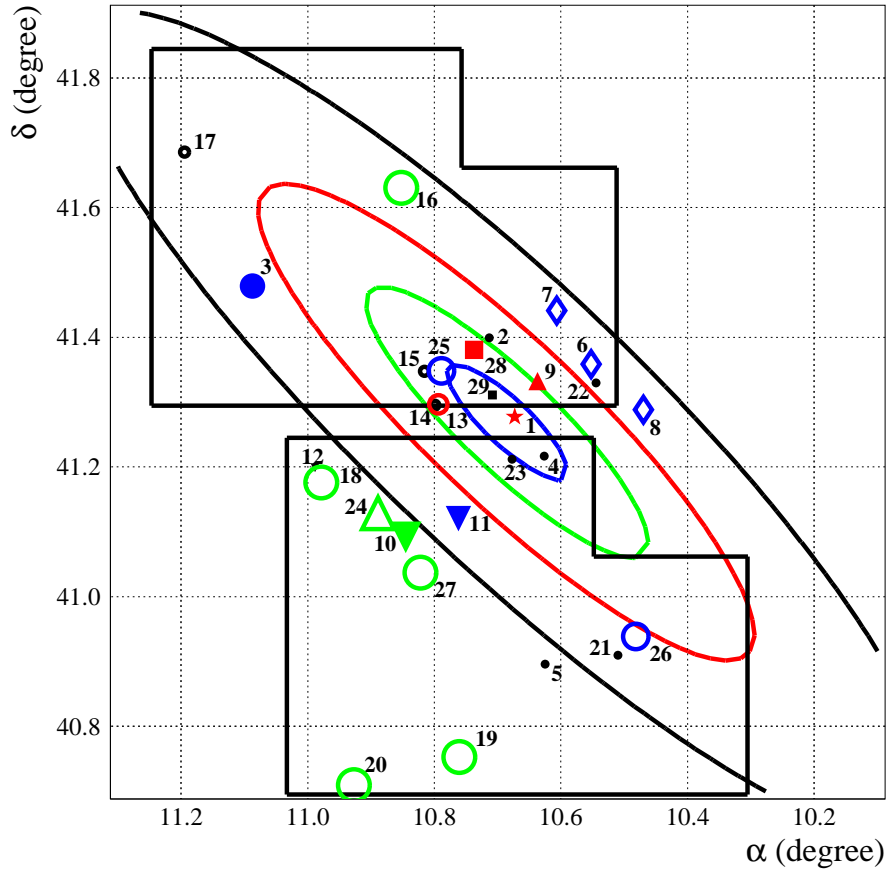}
\caption{Superimposed on iso-density contours of the M31 disc
we report the position of the candidate microlensing events observed
towards M31, indicated according to the numeration
given in Table~\ref{tab:evt_obs}. The symbols refer to the 
collaboration that first reported an event: AGAPE (star), POINT-AGAPE
(filled circles), SLOTT-AGAPE (empty diamonds), MEGA (empty circles), WeCAPP (filled upward triangle),
VATT/Columbia (filled downward triangles), PLAN (filled boxes),
NAINITAL (empty upward triangle). The size of the symbols
is related to the event duration.  Four bins are considered
($t_\mathrm{FWHM}<5~\mathrm{days}$, $5<t_\mathrm{FWHM}<10~\mathrm{days}$,
$10<t_\mathrm{FWHM}<25~\mathrm{days}$ and $t_\mathrm{FWHM}>25~\mathrm{days}$), 
with smaller symbols for shorter duration events (in the
colour version, black, red, blue and green, respectively). Also reported,
the contours of the two fields of view of the 2.5m INT campaign. 
}
\label{fig:m31_evt}    
\end{figure}

\begin{figure}
\includegraphics[width=120mm]{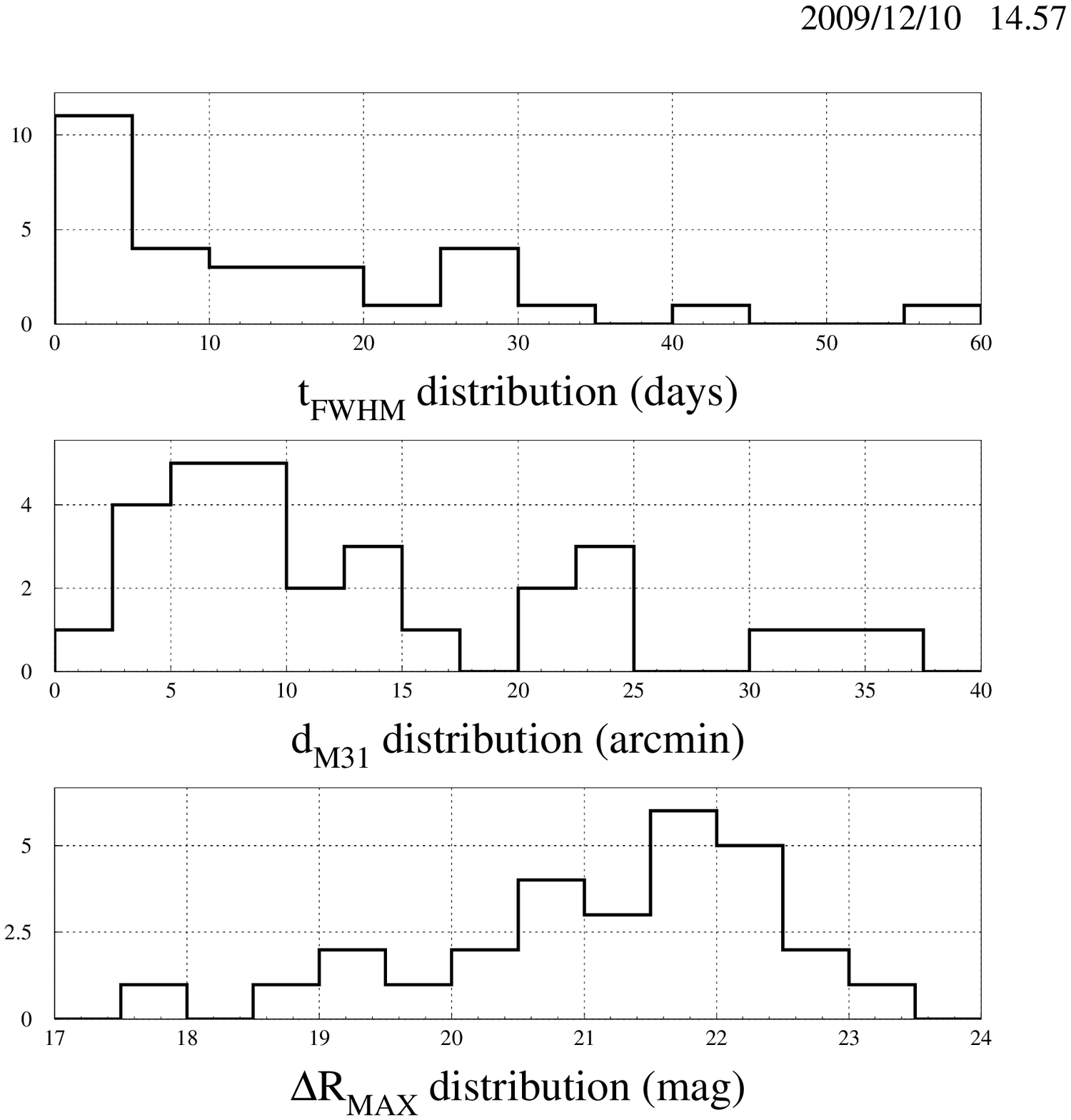}
\caption{For the candidate microlensing events observed
towards M31, Table~\ref{tab:evt_obs}, we report, from top to bottom,
the distribution of the duration, $t_\mathrm{FWHM}$,
of the distance from the M31 centre, $d_\mathrm{M31}$
and of the flux deviation at maximum expressed in term of magnitude, $\Delta R_\mathrm{MAX}$.
}
\label{fig:evt_stat}    
\end{figure}

Microlensing events are distinguished from intrinsic
variable signals for the \emph{shape}, the \emph{achromaticity}
and the \emph{unicity}. As for the shape, most of the events
are expected, and found, to follow the 
symmetric bell-like Paczy\'{n}ski light curve.
Any deviation (finite source effect, parallax,
binary lens/source, \ldots), on the other hand, if explained in a microlensing
context, can only reinforce the microlensing interpretation.
On the other hand, because the background level has not, in general, the same colour than the lensed star,
only the luminosity increase during the microlensing event 
is expected to be achromatic \cite{agape97}.

The very first concern of microlensing campaigns
towards M31, in a regime where
one can not resolve the source stars,
was that of being able to detect 
microlensing events at all.
After more than 10 years now of observational
efforts, there are no doubts that microlensing
events towards M31 have been observed.

The first microlensing candidate events reported
towards M31 have been those presented by the VATT/Columbia
collaboration \cite{crotts96}. 
Although this represented a relevant result,
as this clearly showed, for the first time, 
the possibility of such a detection, these flux variations
(all of them of quite long duration and in fact lacking
both a suitable sampling and a long enough baseline
to robustly probe their unicity)
have not been further discussed in the following
works of the same collaboration, in particular
in \cite{uglesich04}, so we are not going to consider them any longer.

The first convincing microlensing candidate event 
towards M31,  AGAPE-Z1, was then 
observed and characterised by the AGAPE 
collaboration \cite{agape99}. Indeed,
its position and its shape, in particular its short
duration, make of it a quite robust candidate. 
However, and this reason prevented at that time
a sharper conclusion about its microlensing nature,
although the possible intrinsic variable contaminations
were carefully analysed and ruled out,
the lack of a suitable coverage of the flux deviation
in two colours did not allow the achromaticity to be properly tested.

In Table~\ref{tab:evt_obs} we report all 
the microlensing candidate events that have been reported up to now.
A few of them have been detected independently by more
than one collaboration (this holds in particular
for the POINT-AGAPE and the MEGA group who shared
the same set of INT data). The total number of
events sums up to 29. 

In particular, from the analysis of Belokurov et~al. \cite{belokurov05}
we report only the three ``first level'' candidates.
The only new one with respect to the previous POINT-AGAPE
analyses was PA-S16 (this is a POINT-AGAPE internal name only,
in that paper it was simply indicated as ``candidate 1'').
Three ``second level'' candidates were also reported, all of them quite faint
and of rather long duration, $t_\mathrm{FWHM}>30~\mathrm{days}$, and
16 ``third level'' ones. Noteworthy, 11 out of these 16 lie
at more than 12' from the M31 centre.

Most of the sources of M31 microlensing events
are expected to be bright giants stars \cite{agape97},
that should be visible, therefore,
on HST images. Indeed, an identification
of the source has been reported
for a few events. This is extremely
important in particular as it allows
one to break the parameter degeneracy
and to measure the physical timescale $t_\mathrm{E}$.

In Fig.~\ref{fig:m31_evt} we report, 
superimposed on the isodensity contours of the M31 disk,
the position of these candidate events.

In Fig.~\ref{fig:evt_stat} we report, for this
set of observed candidate events, the 
distribution of the duration $t_\mathrm{FWHM}$,
the flux deviation at maximum, expressed in term
of magnitude, $\Delta R_\mathrm{MAX}$, 
and of the distance from the M31 centre.
Looking at these distributions, 
we note that most of the reported events 
have a quite short duration, $t_\mathrm{FWHM}<10~\mathrm{days}$,
and are clustered around the inner M31 region.
These outcomes match well the theoretical expectations (\S~\ref{sec:m31_exp})
and are relevant with respect to the issue of the possible contamination
of variable stars to the microlensing signal.

\begin{figure}
\includegraphics[width=120mm]{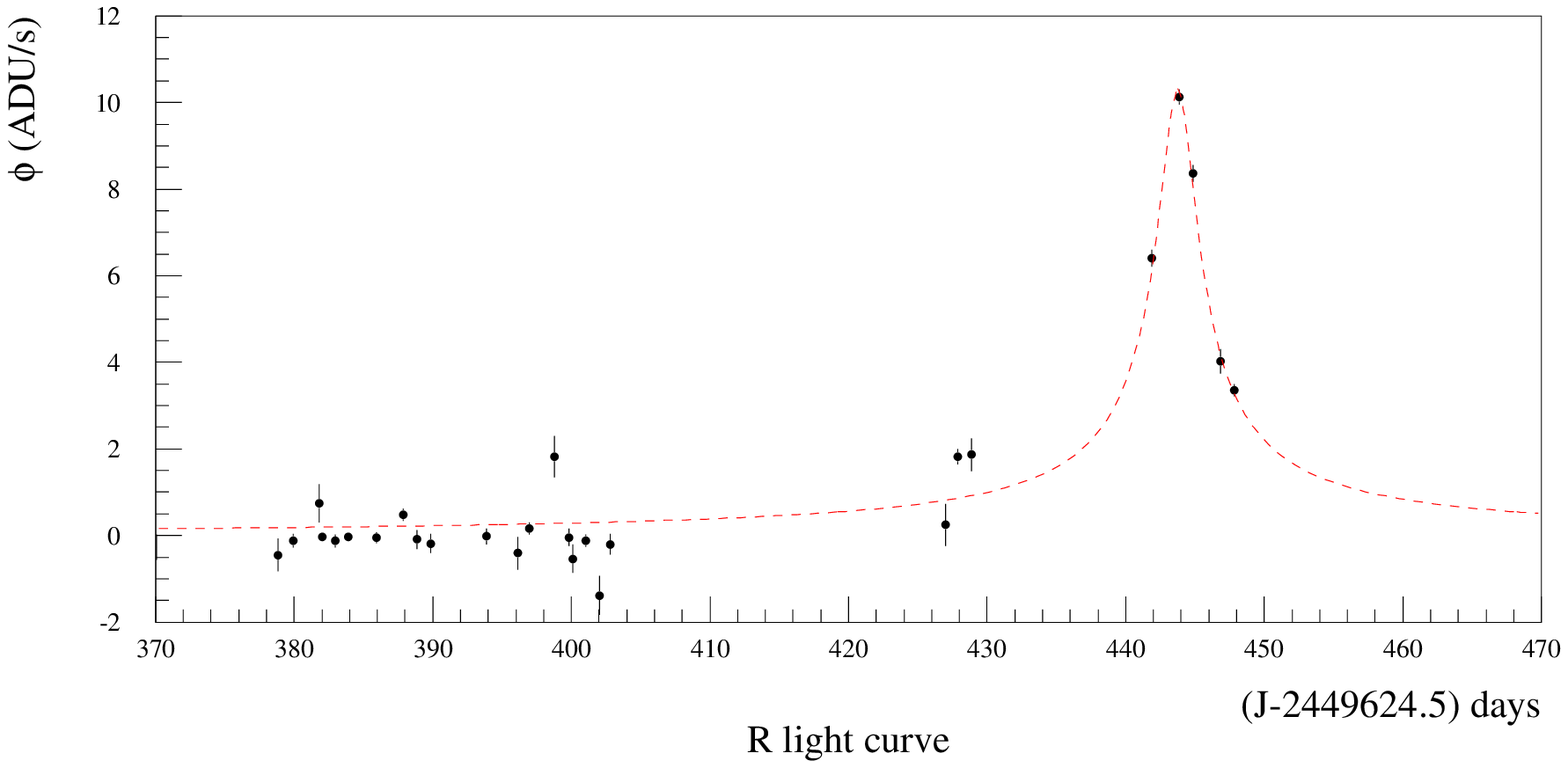}
\includegraphics[width=120mm]{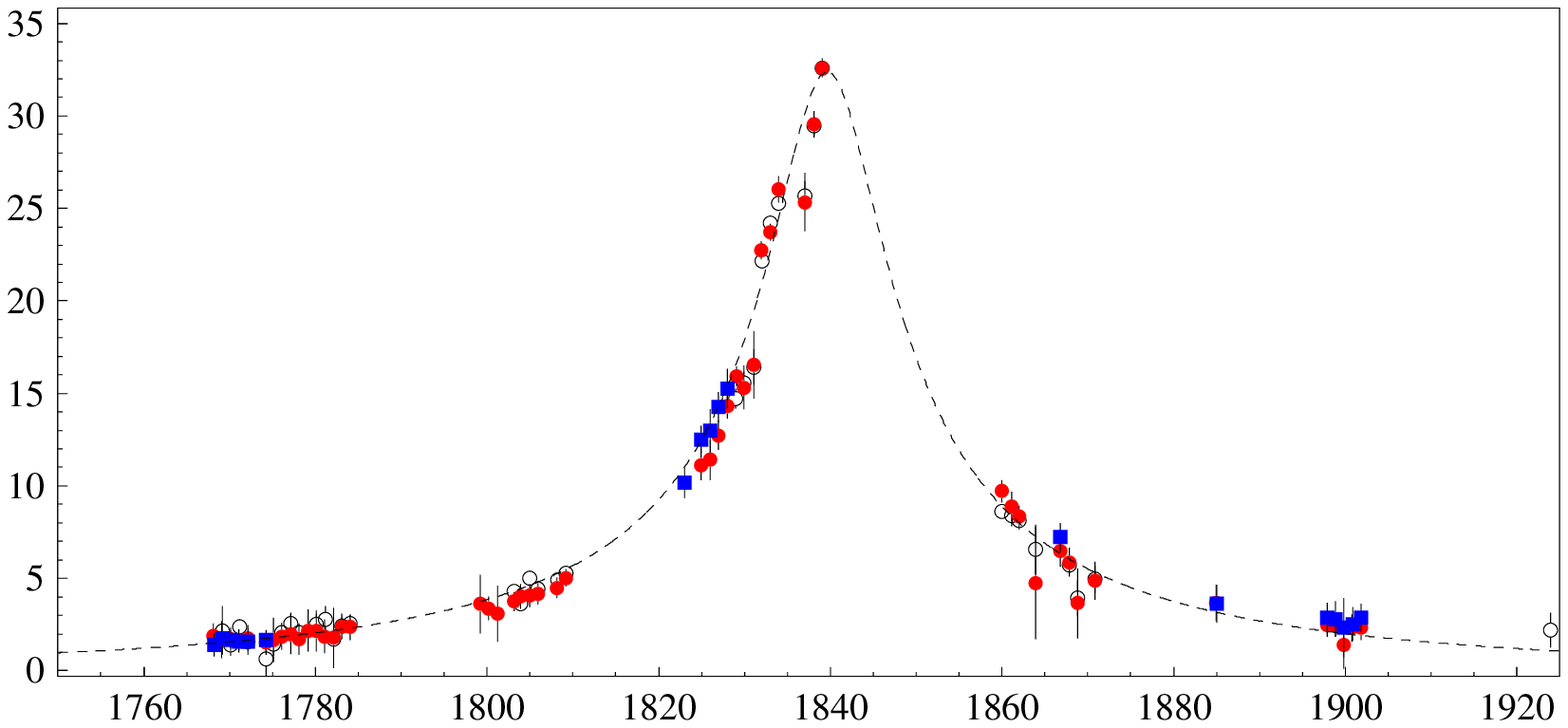}
\includegraphics[width=120mm]{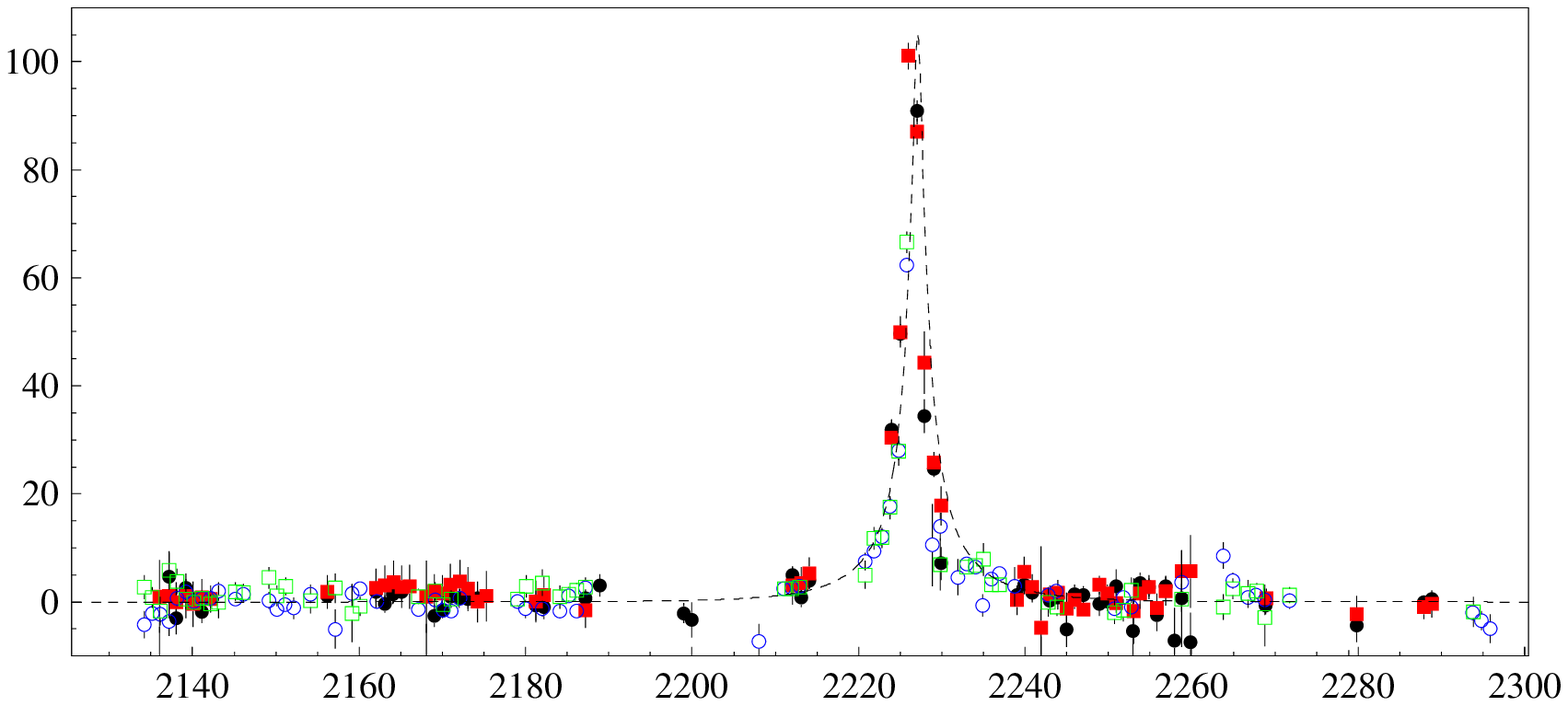}
\caption{The light curves for three pixel lensing (candidate) events observed
towards M31. From top to bottom: AGAPE-Z1 (Figure reproduced from Fig.~3 of \cite{agape99}); 
POINT-AGAPE PA-N2/MEGA ML7 (Figure adapted from \cite{paulin03}); 
POINT-AGAPE S3/WeCAPP GL1 (Figure adapted from \cite{paulin03,wecapp03}). The dashed curves represent
the best  Paczy\'{n}ski light curve fits. The units on the axes (middle
and bottom panels) are the same as in the top panel. Middle panel: 
empty/filled circles and filled boxes are for $g',r'$ and $i'$ band data
respectively (black, red and blue in the colour version).  
Bottom panel: circles and boxes are for $R$ and $I$ band data,
filled and empty symbols for the POINT-AGAPE and WeCAPP data sets.
}
\label{fig:evt_cl}    
\end{figure}

Because of their intrinsically non-repeating nature,
most of the time it is difficult to completely rule out
the possibility of an intrinsic variable contamination
and altogether drop the qualification of \emph{candidate}
for a reported microlensing event. For single events this is actually
possible if the light curve shape deviates, in a microlensing-like way,
from the smooth Paczy\'{n}ski shape, for instance for binary
events. This might be the case for at least one
of the M31 pixel-lensing events, PA-N2.
On the other hand, for a large enough set of events this is possible statistically
if the observed signal characteristics match
the expected ones.

The background noise to the microlensing signal
is given by intrinsically variable stars.
In the pixel-lensing regime many stars
contribute to the flux of each pixel.
Even if not  bright enough to give rise to a detectable signal
over the background noise level, these can add
an additional, non-gaussian, noise to the light curve
(with the extreme case being, as for PA-N1, 
of a bonafide microlensing event superimposed
on the same light curve of a variable star).
This is indeed a relevant issue as for
our ability to select microlensing events at all
(so that it concerns the \emph{efficiency} of a
selection pipeline). 
Second, there might be variable stars masquerading as microlensing events,
and this indeed is the biggest single problem
in the interpretation of microlensing events.
For a typical pixel-lensing campaign one monitors 
thousands of flux variations due to intrinsic variable stars to be compared with
a few (if any) microlensing signals.
Variable stars contaminating the microlensing signal
should reside either in our Galaxy or in M31 itself,
or might be background supernov{\ae}. For 
(unique flux variation) candidate events of short duration and
located near to the M31 centre, 
all of these possible
form of contamination can be, even if not absolutely
ruled out, safely excluded \cite{novati05}.
(For instance, a few flux variations first
claimed to be microlensing candidates,
located in the inner M31 region but with a
quite large timescale \cite{novati02}, have been in a second
moment, once the analysis extended over a longer baseline, 
probed to be due to intrinsically variable   stars \cite{novati03}.)
While this argument can be used statistically
for large enough set of events,
for single events, for which a long enough
flat baseline (a usual demand in any selection
pipeline) can be taken as a strong 
proof against the contamination
of repeating variables, the single one
more dangerous form of contamination
comes from eruptive variables.
This possibility must be carefully considered case by case,
with particular attention to the \emph{shape} and the \emph{colour}
of the candidate event.

Looking back at the distributions shown in Fig.~\ref{fig:evt_stat}
we may indeed wonder whether long duration and faint events,
the most easily to be confused with underlying background variable stars,
can be looked at as outliers of a truly microlensing distribution.

Besides AGAPE-Z1 (Fig.~\ref{fig:evt_cl}, top panel), a few more events deserve some more comments.
PA-N1 \cite{point01} is characteristic in that
its light curve is clearly contaminated by 
a nearby variable. Its short duration,
good sampling and probed achromaticity make of it a robust
microlensing event. Furthermore,
the possible source has been identified
on some HST frame (this identification, however,
has been challenged  by 
Cseresnjes et~al. \cite{mega05_hst}).
PA-N2 \cite{paulin03}/MEGA-ML7 \cite{mega04},
Fig.~\ref{fig:evt_cl} middle panel, is in many ways a peculiar
nonetheless extremely robust microlensing event.
This is because  of its long duration, very large
brightness at maximum amplification
and location, far away from the M31 centre.
In fact, this variation has been
observed and well sampled, for most
of its (long) bump, in three colours,
so that the expected achromaticity
has been extremely well verified.
It is going to be interesting, once the statistics of observed events
will enlarge with new observational campaigns,
to see whether other events with similar characteristics
will be observed or not.
Furthermore, the PA-N2 light curve shows a deviation from the simple
Paczy\'{n}ski shape. This anomaly
has been the object of a thorough analysis of the POINT-AGAPE
collaboration. In particular, An et~al. \cite{an04n2}
probed it to be compatible with a binary lens system.
Because of all of these reasons PA-N2 
looks as a robust microlensing event on its own right.
PA-S3 \cite{paulin03}/GL1 \cite{wecapp03},
besides being, as expected for the ``typical''
M31 pixel-lensing events, short and near the M31
centre,  has been observed both in the INT and the WeCAPP data set.
The joined light curve gives an extremely
convincing bonafide microlensing event (Fig.~\ref{fig:evt_cl}, bottom panel).
We also recall PA-S4 \cite{paulin02}/MEGA-ML11\cite{mega04},
located roughly along the line
of sight of M32, a companion galaxy of M31, which 
is a convincing inter-galactic microlensing event \cite{paulin02}.
Finally we mention PA-S5 \cite{novati05}:
although this flux variation has \emph{not} been selected
by the microlensing selection pipeline
(so that it even lacks the status of \emph{candidate} event),
still, it looks extremely interesting.
Its shape, not to be easily explained
by any intrinsic variable, deviates significantly from
the Paczy\'{n}ski one in a way that is suggestive
of a possible binary lens configuration.
However, the poor sampling along the bump
did not allow its full characterization.
A further reason of interest is its position,
about $20'$ away from the M31 centre.

\subsubsection{Looking for compact halo objects (and self-lensing events)} 
\label{sec:m31_macho}

Once acknowledged the possibility
to detect and characterize microlensing events
in the pixel-lensing regime towards M31,
the leading scientific question that has been
addressed is the search for dark matter in form of compact
halo objects. As already outlined, a main
problem is the ability to distinguish MACHO lensing events
from self-lensing ones. In fact,
as we detail below, the main results reported up to now
are in disagreement on the MACHO halo content,
and this can be traced back mainly to this issue.

The first attempt to draw conclusions on the MACHO content
towards M31 has been carried out by Uglesich et~al. \cite{uglesich04}, 
who concluded for an evidence of a MACHO signal.
The next analyses, from the POINT-AGAPE and 
the MEGA collaboration, presented their results 
with more detailed and reliable efficiency analyses,
an essential step to meaningfully compare
the observed and the expected signal.

POINT-AGAPE and MEGA made use of the same
INT data set (although MEGA considered
a fourth year of data not included in the POINT-AGAPE
analysis). 

POINT-AGAPE reported an \emph{evidence}
for a MACHO signal towards M31 \cite{novati05}.
In particular they have evaluated a \emph{lower} limit
for the halo fraction in form of MACHOs, $f$,
of about 20\% in the mass range $0.1-1~\mathrm{M}_\odot$.
On the other hand, MEGA concluded \cite{mega06} that their observed
rate was consistent with the expected self-lensing signal.
In particular they ruled out a MACHO halo fraction,
for $0.5~\mathrm{M}_\odot$ compact halo objects, larger than 30\%.

POINT-AGAPE restricted the search of microlensing events to short duration,
$t_\mathrm{FWHM}<25~\mathrm{days}$, and bright,
$\Delta R_\mathrm{MAX}<21$, flux variations.
With a thorough discussion to exclude the
contamination of intrinsic variable objects
they presented 5 microlensing candidate events
upon which they based their following 
analysis\footnote{A sixth candidate events, PA-S4,
was not included as acknowledged to be,
more likely, an intergalactic M31-M32 event.}.
Without any a priori cut in the event parameter
space, MEGA presented 14 microlensing candidate events.
All the additional events with respect to POINT-AGAPE
can be explained because of the enlarged parameter space
and the extended overall baseline.
Besides, MEGA failed to report the detection
of 2 of the POINT-AGAPE candidates
(plus a third one detected nearby the M31 centre
in  a region MEGA excluded from his analysis).

To evaluate the expected signal,
POINT-AGAPE developed a full Monte Carlo simulation
completed by an efficiency analysis 
where the events selected within the Monte Carlo
were injected in the data and submitted to the analysis pipeline.
MEGA evaluated the microlensing rate 
taking into account the pipeline detection efficiency.
As a result, for their fiducial model,
to be specific we are going hereafter to consider
MACHOs of $0.5~\mathrm{M}_\odot$ and fix $f=20\%$,
POINT-AGAPE reported an expected self-lensing signal
of about 0.8 events to be compared with 1.4 MACHO events,
MEGA of 14 versus 6.2 events, respectively. Besides the overall numbers,
that can not be directly compared because
of the different regions in the event parameter space
explored, a striking difference is 
in the expected \emph{ratio} of self-lensing over MACHO lensing.

In order to evaluate the probability function
for $f$, POINT-AGAPE have taken a specific care
in order to include the information on the spatial
distribution of the observed events.
In fact, the position of PA-N2 , \S~\ref{sec:m31_evt}, well away
from the M31 inner region where self lensing
is expected to be relatively small with respect
to the would be MACHO signal, turned out
to be essential to conclude  on the ``evidence''
for the MACHO signal. On the other hand, a main weakness in the POINT-AGAPE analysis
is the small statistics. This holds in particular
because the reported result heavily depends on a single,
though extremely robust by itself, microlensing candidate (PA-N2).
Furthermore, MEGA argued that the M31 model 
used by POINT-AGAPE was bound to unduly \emph{underestimate}
the expected self-lensing signal even though 
POINT-AGAPE claimed to have paid attention, for his fiducial
model, to actually consider \emph{stellar} lenses only 
to account for self lensing. Besides, POINT-AGAPE
have tested their results also
for more massive luminous models, \S~\ref{sec:m31_mod},
always finding evidence for a MACHO signal.

Although the analysis on the \emph{number}
of observed candidates versus expected self-lensing events
clearly points towards the self-lensing explanation
of the observed rate, a weakness in the MEGA
analysis can be found in that they seem
to underestimate the role played by the characteristics
of the observed events. In particular,
the spatial distribution of the events,
both for a signal of asymmetry and because
quite a large fraction of them lie
far from the innermost M31 region,
seems  indeed, as in fact also pointed out by MEGA
(see in particular their Fig.~18),
to favour MACHO lensing over self lensing.
(Besides, in their previous analysis, de~Jong et~al. \cite{mega04},
the MEGA collaboration had preliminarly concluded
that ``the spatial distribution of candidate
microlenses is suggestive of the presence of
a microlensing halo'', and more specifically
``dark'' halo). The spatial distribution, 
and a few more issues, have been also analysed 
by Ingrosso et~al. \cite{ingrosso06,ingrosso07}
who concluded that self lensing can hardly
explain all of the MEGA microlensing candidates.

These analyses clearly have left the MACHO issue,
as for the line of sight towards M31, still open.
They also show the extent to which a correct  modelling
of M31, in particular of its luminous components,
is extremely relevant. In fact, this
should be taken as an opportunity of the 
need of a better astrophysical
understanding both of the expected signal and
of the observed events.

Along this direction the WeCAPP collaboration
made a relevant contribution
with an extremely detailed analysis
of the PA-S3/GL1 event \cite{arno08}.
The joint analysis of INT and WeCAPP data
allowed them to strongly constrain this event.
Furthermore, a specific analysis
on the relevance of the finite source effect,
considering in particular the extremely large brightness of the event,
and carefully including an analysis of all the event characteristics,
allowed the authors to conclude on
the much more likely MACHO, rather than 
stellar, nature of the lens. It is also noteworthy 
that this result is reached although, \S~\ref{sec:m31_evt}, this event
lies at very short distance from the M31 centre,
where self-lensing signal is expected to be quite large
with respect to MACHO lensing.
Finally, it is worth stressing the methodological importance
of such a joint analysis of different data sets
(even reduced following different photometry schemes)
that in particular allowed to robustly confirm the microlensing
nature of this flux variation.
This is a quite obvious outcome for Galactic bulge searches
and it clearly suggests the way to be followed
also for M31 analyses.

\subsection{The hunt for extra-solar (extra-galactic) planets} 
\label{sec:m31_eso}

Beyond the search for MACHOs microlensing is, together with other techniques
(see e.g. the review in \cite{perryman05}),
a suitable tool for the detection of extra-solar planets
\cite{maopacz91,gould92}.
The microlensing signal for a planetary system
is that of a binary lens with extremely small mass ratio, $q$,
and it shows itself as a short duration perturbation
of the smooth single lens light curve. Only
of order of ten planets have been detected, towards the Galactic bulge,
using microlensing (e.g. \cite{gould09}) against
a few hundreds with other methods (mainly, radial velocity
and transit). However, microlensing has various
advantages over other methods \cite{dominik08,beaulieu08}.
In particular, microlensing is the only available
tool sensitive to extra-solar planets at large distance from the
solar system, up to extra-galactic distances. 

Covone et~al. \cite{covone00} and Baltz \& Gondolo \cite{baltz01}
have been the first to discuss the possibility
to detect extra-solar planets in M31 with pixel lensing.
More recently, Chung et~al. \cite{chung06},
addressed this issue within the specific experimental set up
of the ANGSTROM project
discussing, in particular, the efficiency for detecting
planets of Jupiter-mass  within the lensing zone
using a full network of telescopes.
Ingrosso et~al. \cite{ingrosso09} have analysed the same issue
with the additional bonus of
using a Monte Carlo approach spanning both
the parameter space of the lens-planet system
as well as that of the underlying lensing events. In particular,
they have pointed out that a $\sim 6$-Jupiter mass planet in M31 might already
have been detected. In fact, the anomaly of the POINT-AGAPE
PA-N2 candidate event had already been discussed
to be compatible with a binary lens system with
an extremely small mass ratio \cite{an04n2}.
As a caveat for this exciting outcome we recall the extremely
large uncertainties in the lens mass determination
(and therefore on that of its companion).
Whatever be the case with PA-N2,  from the above
considerations it clearly appears that the detection
of planets in M31 is already a reachable objective.
To this purpose, however, a similar strategy,
with the caveats suggested by all the peculiarities
of the pixel-lensing regime, as that
followed for microlensing planet searches towards
the Galactic bulge, with survey and follow up
all around the world, is needed. 

Finally, we mention a relevant byproduct
of M31 pixel lensing searches, as of all 
microlensing analyses, namely the study
of variable stars. This is both interesting in itself
but also as a way to better understand
the microlensing signal.  We have already mentioned
a relevant outcome from the thorough analysis of the POINT-AGAPE catalogue \cite{an04}.
We also recall the specific analyses of Nov{\ae} by
POINT-AGAPE \cite{novae04,novae06}, 
the AGAPE \cite{ansari04} and WeCAPP M31 variable star catalogues \cite{fliri06},
and the Nainital analyses \cite{joshi03,joshi04}.

\section{Pixel lensing beyond the Local Group} \label{sec:m31_beyond}

The pixel lensing technique allows 
the search for (stellar) microlensing events
to be extended beyond the limit of the Local Group.
This opportunity is extremely challenging,
also because, at such large distance, one fully enters
the ``spike'' regime, \S~\ref{sec:pl_theory}. 
Pixel lensing is only moving his first steps 
in this direction, nonetheless it already clearly probed
to be a viable tool of analysis.

A first possible target was soon identified to be
the giant elliptical galaxy M87 at the centre of the Virgo Cluster.
In \cite{gould95b} Gould addressed this problem
and made a detailed proposal for a WFPC2/HST campaign
with this purpose. Baltz et~al \cite{baltz04}
carried out this ambitious programme
with a campaign lasting 30 days. In particular, they reported
the detection of seven variable sources
among those they identified one viable
microlensing candidate consistent
with a dark matter halo mass fraction of about 20\%
of microlensing objects for both M87 and the Virgo cluster.

A few other targets have also been proposed: the cluster A2152 in \cite{totani03}
and A2218 and A370 in \cite{tuntsov04}. More recently,
de~Jong et~al. \cite{dejong08} carried out a microlensing pilot campaign 
looking at Centaurus A using the ESO/MPG 2.2m telescope,
showing in particular the feasibility of the project.

\section{Conclusion}

Pixel lensing is stellar microlensing of unresolved sources
(a situation characterized, as we have outlined,
by the fact that the photon noise is dominated
by that of the background level).
It allows the realm of microlensing to be extended to
distant targets, in the Local Group and beyond.
In this review we have focused on the main
observational results obtained up to now.
The principal target for observations
has been our nearby galaxy, M31.

Pixel lensing has probed to be able to confidently detect
and robustly characterize microlensing events.
The original microlensing
motivation, the search for the 
(dark matter) compact halo objects (MACHOs) signal,
is still an open issue. In the meantime
it has become clear the importance of 
a correct understanding of self lensing,
for which the lens belongs to some luminous population,
both as a background signal
to MACHO lensing and as an opportunity
in itself for the study of the luminous
lensing components.  

The detection of about 30 microlensing candidate events
has been reported towards M31. In particular, out
of a complete analysis on the same INT data set,
the POINT-AGAPE collaboration concluded
for an evidence of a MACHO signal
whereas the MEGA collaboration rather 
found its detected signal to be compatible
with the expected self-lensing one.
On the other hand, the careful analysis of an extremely
well sampled event (with two different data sets)
allowed the WeCAPP collaboration to conclude
on the more likely MACHO rather than  stellar
nature for the lens of a bright event
detected in the central M31 region.

These results motivate to further carry on
both the theoretical and the observational efforts.
In fact, pixel lensing is now entering
a new phase of maturity, rich in opportunities not to be missed,
with in particular the need for a deeper understanding of the 
microlensing signal from an astrophysical point of view.
The increase of the understanding
of the lensing signal with the ongoing
campaign (ANGSTROM and PLAN), 
the effort to establish a real time analysis
to be used for a network of telescopes (ANGSTROM),
as well as the significant increase
also in the number of events expected with the PAandromeda project,
with the perspective of an invaluable full coverage of M31, 
are the next to come and essential steps towards this purpose.

Beyond the search for dark matter in form of compact halo
objects, and more generally the study of the
luminous M31 lens populations with self lensing,
M31 pixel lensing is beginning to face also
a new challenging purpose, still already within
the reach of the present technology, the search
for extra-solar, extra-galactic, planets. This is 
an additional (if needed) strong motivation for M31 pixel lensing searches.

\vspace{1cm}
\emph{Note added in proof}\\ 
The WeCAPP collaboration  (A. Riffeser and S. Seitz, private communication) 
is currently completing the final analysis of their 11-years
campaign (Riffeser et al., 2010, in preparation
and Koppenh\"ofer et al., 2010, in preparation). 
They report the detection of 10 microlensing events (all of them
with very short duration and specifically, 8 out of 10
with $t_\mathrm{FWHM}<5~\mathrm{days}$). Their preliminar
results on the expected rate indicate
that the self-lensing signal alone is not
sufficient to explain all of the observed events.

\begin{acknowledgements}
It is a pleasure to thank the editors
of the present volume, and in particular
Ph.~Jetzer, for giving me the opportunity
to write this review.
I would like to thank the AGAPE group, in particular 
Y.~Giraud-H\`eraud, J.~Kaplan, M.~Cr\'ez\'e and P.~Baillon,
for introducing and carrying me on through
this fascinating subject in the course of several years.
I acknowledge support for this work
by the Italian Space Agency (ASI)
and by the ``Istituto Internazionale per gli 
Alti Studi Scientifici'' (IIASS).
\end{acknowledgements}

\bibliographystyle{spphys}       % APS-like style for physics
\bibliography{biblio}   % name your BibTeX data base

\begin{thebibliography}{10}
\providecommand{\url}[1]{{#1}}
\providecommand{\urlprefix}{URL }
\expandafter\ifx\csname urlstyle\endcsname\relax
  \providecommand{\doi}[1]{DOI \discretionary{}{}{}#1}\else
  \providecommand{\doi}{DOI \discretionary{}{}{}\begingroup
  \urlstyle{rm}\Url}\fi

\bibitem{pacz86}
B.~Paczy\'{n}ski, \apj,~ \textbf{304}, 1 (1986)

\bibitem{moniez09}
M.~{Moniez}, {Microlensing towards the galactic center and LMC/SMC},
\newblock This volume

\bibitem{dominik09b}
M.~{Dominik}, {Microlensing and planet detection},
\newblock This volume

\bibitem{macho00}
C.~{Alcock}, R.A. {Allsman}, D.R. {Alves}, {et~al.}, \apj,~
  \textbf{542}, 281 (2000)

\bibitem{bennett05}
D.P. {Bennett}, \apj ~\textbf{633}, 906 (2005)

\bibitem{eros07}
P.~{Tisserand}, L.~{Le Guillou}, C.~{Afonso}, {et~al.} \aap,~ 
\textbf{469}, 387 (2007)

\bibitem{ogle09}
L.~{Wyrzykowski}, S.~{Kozlowski}, J.~{Skowron}, {et~al.} \mnras,~
\textbf{397}, 1228 (2009)


\bibitem{wambsganss09}
J.~{Wambsganss}, {Extragalctic microlensing},
\newblock This volume

\bibitem{pacz96}
B.~{Paczynski}, \araa,~ \textbf{34}, 419 (1996)

\bibitem{roulet97}
E.~{Roulet}, S.~{Mollerach}, \physrep,~ \textbf{279}, 67 (1997)

\bibitem{wambsganss06}
J.~{Wambsganss}, \emph{{Gravitational Microlensing}} 
\newblock in Gravitational Lensing: Strong, Weak and Micro, 
Saas-Fee Advanced Courses, Volume 33.~Springer-Verlag Berlin Heidelberg, p.~453 
(2006)

\bibitem{gould96}
A.~Gould, \apj,~ \textbf{470}, 201 (1996)

\bibitem{popowski05}
P.~{Popowski}, K.~{Griest}, C.L. {Thomas}, {et~al.}  \apj,~ \textbf{631}, 879 (2005)

\bibitem{hamadache06}
C.~{Hamadache}, L.~{Le Guillou}, P.~{Tisserand}, {et~al.} \aap,~
  \textbf{454}, 185 (2006)

\bibitem{sumi06}
T.~{Sumi}, P.R. {Wo{\'z}niak}, A.~{Udalski}, {et~al.}, \apj,~ \textbf{636}, 240 (2006)

\bibitem{wozniak_pacz87}
P.~{Wozniak}, B.~{Paczynski}, \apj,~ \textbf{487}, 55 (1997)

\bibitem{gondolo99}
P.~{Gondolo}, \apjl,~ \textbf{510}, L29 (1999)

\bibitem{baltz00}
E.A. {Baltz}, J.~{Silk}, \apj,~ \textbf{530}, 578 (2000)

\bibitem{dominik09a}
M.~{Dominik}, \mnras,~ \textbf{393}, 816 (2009)


\bibitem{kerins01}
E.~{Kerins}, B.J. {Carr}, N.W. {Evans}, {et~al.}, \mnras,~ \textbf{323},
  13 (2001)

\bibitem{melchior98}
A.~{Melchior}, C.~{Afonso}, R.~{Ansari}, et~al., \aap
  ~\textbf{339}, 658 (1998)

\bibitem{melchior99}
A.~{Melchior}, C.~{Afonso}, R.~{Ansari},  et~al., \aaps ~\textbf{134}, 377 (1999)


\bibitem{crotts92}
A.P.S. {Crotts}, \apj,~ \textbf{399}, L43 (1992).

\bibitem{agape93}
P.~{Baillon}, A.~{Bouquet}, Y.~{Giraud-Heraud}, J.~{Kaplan}, \aap,~ \textbf{277},
  1 (1993)

\bibitem{tomaney96}
A.B. {Tomaney}, A.P.S. {Crotts}, \aj,~ \textbf{112}, 2872 (1996)

\bibitem{agape97}
R.~Ansari, M.~Auri\`ere, P.~Baillon, {et~al.}, 
\aap,~  \textbf{324}, 843 (1997)


\bibitem{jetzer94}
P.~{Jetzer}, \aap,~ \textbf{286}, 426 (1994)

\bibitem{colley95}
W.N. {Colley}, \aj,~ \textbf{109}, 440 (1995)


\bibitem{han96}
C.~{Han}, \apj,~ \textbf{472}, 108 (1996)

\bibitem{gyuk_crotts00}
G.~{Gyuk}, A.~{Crotts}, \apj,~ \textbf{535}, 621 (2000)


\bibitem{baltz03}
E.A. {Baltz}, G.~{Gyuk}, A.~{Crotts}, \apj,~ \textbf{582}, 30 (2003)

\bibitem{baltz05}
E.A. {Baltz}, \apj,~ \textbf{624}, 168 (2005)

\bibitem{arno06}
A.~{Riffeser}, J.~{Fliri}, S.~{Seitz}, R.~{Bender}, \apjs ~\textbf{163}, 225
  (2006)


\bibitem{novati05}
S.~{Calchi Novati}, S.~{Paulin-Henriksson}, J.~{An}, {et~al.}, \aap,~
  \textbf{443}, 911 (2005)

\bibitem{novati09}
S.~{Calchi Novati}, V.~{Bozza}, F.~{DePaolis}, {et~al.}, 
\apj,~ \textbf{695}, 442 (2009)


\bibitem{crotts96}
A.P.S. {Crotts}, A.B. {Tomaney}, \apjl,~ \textbf{473}, L87 (1996)

\bibitem{uglesich04}
R.R. {Uglesich}, A.P.S. {Crotts}, E.A. {Baltz} {et~al.}, \apj,~ 
\textbf{612}, 877 (2004)

\bibitem{mega04}
J.T.A. {de Jong}, K.~{Kuijken}, A.P.S. {Crotts} {et~al.}, 
\aap,~ \textbf{417}, 461 (2004)

\bibitem{mega06}
J.T.A. {de Jong}, L.M. {Widrow}, P.~{Cseresnjes}, {et~al.},
\aap,~ \textbf{446}, 855 (2006)


\bibitem{agape99}
R.~Ansari, M.~Auri\`ere, P.~Baillon, {et~al.}, 
\aap,~ \textbf{344}, L49 (1999)

\bibitem{novati02}
S.~{Calchi Novati}, G.~{Iovane}, A.A. {Marino}, {et~al.}, 
\aap,~  \textbf{381}, 848 (2002)

\bibitem{novati03}
S.~{Calchi Novati}, P.~{Jetzer}, G.~{Scarpetta}, {et~al.}, 
\aap,~ \textbf{405}, 851 (2003)

\bibitem{joshi05}
Y.C. {Joshi}, A.K. {Pandey}, D.~{Narasimha}, R.~{Sagar}, 
\aap,~ \textbf{433}, 787  (2005)

\bibitem{point01}
M.~{Auri{\`e}re}, P.~{Baillon}, A.~{Bouquet}, {et~al.}, 
\apjl,~ \textbf{553}, L137 (2001)

\bibitem{paulin03}
S.~{Paulin-Henriksson}, P.~{Baillon}, A.~{Bouquet}, {et~al.}, 
\aap,~ \textbf{405}, 15 (2003)

\bibitem{belokurov05}
V.~{Belokurov}, J.~{An}, N.W. {Evans}, {et~al.}, \mnras,~
  \textbf{357}, 17 (2005)

\bibitem{wecapp01}
A.~{Riffeser}, J.~{Fliri}, C.A. {G{\"o}ssl},  et~al., \aap,
~\textbf{379}, 362 (2001)

\bibitem{wecapp03}
A.~{Riffeser}, J.~{Fliri}, R.~{Bender}, S.~{Seitz}, C.A. {G{\" o}ssl}, 
\apjl,~\textbf{599}, L17 (2003)

\bibitem{paulin02}
S.~{Paulin-Henriksson}, P.~{Baillon}, A.~{Bouquet}, {et~al.},
\apjl,~ \textbf{576}, L121 (2002)

\bibitem{kerins06}
E.~{Kerins}, M.J. {Darnley}, J.P. {Duke}, {et~al.}, 
\mnras,~ \textbf{365}, 1099 (2006)

\bibitem{novati07}
S.~{Calchi Novati}, G.~{Covone}, F.~{de Paolis}, {et~al.}, 
\aap,~ \textbf{469}, 115 (2007)

\bibitem{darnley07}
M.J. {Darnley}, E.~{Kerins}, A.~{Newsam}, {et~al.}, \apjl
~\textbf{661}, L45 (2007)

\bibitem{kim07}
D.~{Kim}, S.J. {Chung}, M.J. {Darnley}, {et~al.}, \apj ~\textbf{666}, 236 (2007)

\bibitem{mcconnachie05}
A.W. {McConnachie}, M.J. {Irwin}, A.M.N. {Ferguson},  et~al., 
\mnras ~\textbf{356}, 979 (2005)

\bibitem{tammann08}
G.A. {Tammann}, A.~{Sandage}, B.~{Reindl}, \apj ~\textbf{679}, 52 (2008)

\bibitem{walterbos87}
R.A.M. {Walterbos}, R.C. {Kennicutt}, Jr., \aaps ~\textbf{69}, 311 (1987)

\bibitem{kent89}
S.M. {Kent}, \aj ~\textbf{97}, 1614 (1989)

\bibitem{chemin09}
L.~{Chemin}, C.~{Carignan}, T.~{Foster}, \apj~in press, arXiv:0909.3846


\bibitem{geehan06}
J.J. {Geehan}, M.A. {Fardal}, A.~{Babul}, P.~{Guhathakurta}, \mnras
  ~\textbf{366}, 996 (2006)

\bibitem{widrow03}
L.M. {Widrow}, K.M. {Perrett}, S.H. {Suyu}, \apj ~\textbf{588}, 311 (2003)

\bibitem{tamm07b}
E.~{Tempel}, A.~{Tamm}, P.~{Tenjes}, arXiv:0707.4374

\bibitem{klypin02}
A.~{Klypin}, H.~{Zhao}, R.S. {Somerville}, \apj ~\textbf{573}, 597 (2002)

\bibitem{montalto09}
M.~{Montalto}, S.~{Seitz}, A.~{Riffeser}, et~al., arXiv:0907.0669


\bibitem{kent83}
S.M. {Kent}, \apj ~\textbf{266}, 562 (1983)

\bibitem{kent86}
S.M. {Kent}, \aj ~\textbf{91}, 1301 (1986)

\bibitem{kent87}
S.M. {Kent}, \aj ~\textbf{94}, 306 (1987)

\bibitem{widrow05}
L.M. {Widrow}, J.~{Dubinski}, \apj ~\textbf{631}, 838 (2005)

\bibitem{athanassoula06}
E.~{Athanassoula}, R.L. {Beaton}, \mnras ~\textbf{370}, 1499 (2006)

\bibitem{beaton07}
R.L. {Beaton}, S.R. {Majewski}, P.~{Guhathakurta},  et~al., \apjl
  ~\textbf{658}, L91 (2007)

\bibitem{saglia09}
R.P. {Saglia}, M.~{Fabricius}, R.~{Bender}, et~al. 
\aap~in press, arXiv:0910.5590

\bibitem{font08}
A.S. {Font}, K.V. {Johnston}, A.M.N. {Ferguson}, et~al., \apj ~\textbf{673}, 215
  (2008)

\bibitem{tanaka09}
M.~{Tanaka}, M.~{Chiba}, Y.~{Komiyama}, P.~{Guhathakurta}, J.S. {Kalirai},
  M.~{Iye}, arXiv:0908.0245

\bibitem{ferguson02}
A.M.N. {Ferguson}, M.J. {Irwin}, R.A. {Ibata}, G.F. {Lewis}, N.R. {Tanvir}, \aj
  ~\textbf{124}, 1452 (2002)

\bibitem{richardson08}
J.C. {Richardson}, A.M.N. {Ferguson}, R.A. {Johnson}, et~al., 
\aj ~\textbf{135}, 1998 (2008)

\bibitem{rich95}
R.M. {Rich}, K.J. {Mighell}, \apj ~\textbf{439}, 145 (1995)


\bibitem{stephens03}
A.W. {Stephens}, J.A. {Frogel}, D.L. {DePoy}, et~al., \aj ~\textbf{125}, 2473
  (2003)

\bibitem{sarajedini05}
A.~{Sarajedini}, P.~{Jablonka}, \aj ~\textbf{130}, 1627 (2005)

\bibitem{olsen06}
K.A.G. {Olsen}, R.D. {Blum}, A.W. {Stephens}, T.J. {Davidge}, P.~{Massey}, S.E.
  {Strom}, F.~{Rigaut}, \aj ~\textbf{132}, 271 (2006)


\bibitem{yin09}
J.~{Yin}, J.L. {Hou}, N.~{Prantzos}, S.~{Boissier}, R.X. {Chang}, S.Y. {Shen},
  B.~{Zhang}, \aap ~\textbf{505}, 497 (2009)

\bibitem{ballero07}
S.K. {Ballero}, P.~{Kroupa}, F.~{Matteucci}, \aap ~\textbf{467}, 117 (2007)

\bibitem{vdm08}
R.P. {van der Marel}, P.~{Guhathakurta}, \apj ~\textbf{678}, 187 (2008)

\bibitem{braun91}
R.~{Braun}, \apj ~\textbf{372}, 54 (1991)

\bibitem{nfw96}
J.F. {Navarro}, C.S. {Frenk}, S.D.M. {White}, \apj ~\textbf{462}, 563 (1996)

\bibitem{nfw97}
J.F. {Navarro}, C.S. {Frenk}, S.D.M. {White}, \apj ~\textbf{490}, 493 (1997)

\bibitem{merritt06}
D.~{Merritt}, A.W. {Graham}, B.~{Moore}, J.~{Diemand}, B.~{Terzi{\'c}}, \aj
  ~\textbf{132}, 2685 (2006)

\bibitem{novati08}
S.~{Calchi Novati}, F.~{de Luca}, P.~{Jetzer}, L.~{Mancini}, G.~{Scarpetta},
  \aap ~\textbf{480}, 723 (2008)

\bibitem{an04}
J.H. {An}, N.W. {Evans}, P.~{Hewett}, et~al., \mnras ~\textbf{351}, 1071 (2004)


\bibitem{gyuk00}
G.~{Gyuk}, N.~{Dalal}, K.~{Griest}, \apj ~\textbf{535}, 90 (2000)


\bibitem{jetzer02}
P.~{Jetzer}, L.~{Mancini}, G.~{Scarpetta}, \aap ~\textbf{393}, 129 (2002)

\bibitem{mancini04}
L.~{Mancini}, S.~{Calchi Novati}, P.~{Jetzer}, G.~{Scarpetta}, \aap
  ~\textbf{427}, 61 (2004)

\bibitem{novati06}
S.~{Calchi Novati}, F.~{De Luca}, P.~{Jetzer}, G.~{Scarpetta}, \aap
  ~\textbf{459}, 407 (2006)

\bibitem{novati09b}
S.~{Calchi Novati}, L.~{Mancini}, G.~{Scarpetta}, {\L}.~{Wyrzykowski}, \mnras
  ~\textbf{400}, 1625 (2009)

\bibitem{witt_mao_94}
H.J. {Witt}, S.~{Mao}, \apj ~\textbf{430}, 505 (1994)

\bibitem{arno08}
A.~{Riffeser}, S.~{Seitz}, R.~{Bender}, \apj ~\textbf{684}, 1093 (2008)

\bibitem{gould94}
A.~{Gould}, \apjl ~\textbf{421}, L71 (1994)

\bibitem{hangould96b}
C.~{Han}, A.~{Gould}, \apj ~\textbf{473}, 230 (1996)

\bibitem{mega05_hst}
P.~{Cseresnjes}, A.P.S. {Crotts}, J.T.A. {de Jong}, {et~al.}, 
\apjl,~ \textbf{633}, L105 (2005)



\bibitem{an04n2}
J.H. {An}, N.W. {Evans}, E.~{Kerins}, {et~al.} \apj,~ \textbf{601}, 845 (2004)


\bibitem{ingrosso06}
G.~{Ingrosso}, S.~{Calchi Novati}, F.~{de Paolis}, et~al., 
\aap,~ \textbf{445}, 375 (2006)

\bibitem{ingrosso07}
G.~{Ingrosso}, S.~{Calchi Novati}, F.~{de Paolis}, et~al., 
\aap,~ \textbf{462}, 895 (2007)


\bibitem{perryman05}
M.~{Perryman}, O.~{Hainaut}, D.~{Dravins}, et~al., arXiv:astro-ph/0506163
%{Report by the ESA-ESO Working Group on  Extra-Solar Planets}, 


\bibitem{maopacz91}
S.~{Mao}, B.~{Paczynski}, \apjl ~\textbf{374}, L37 (1991)

\bibitem{gould92}
A.~{Gould}, A.~{Loeb}, \apj ~\textbf{396}, 104 (1992)

\bibitem{gould09}
A.~{Gould}, \emph{ASPC}, vol. 403, ed.
  by {K.~Z.~Stanek} (2009), vol. 403, p. 86, arXiv:0803.4324

\bibitem{dominik08}
M.~{Dominik}, U.G. {Jorgensen}, K.~{Horne}, et~al., arXiv:0808.0004

\bibitem{beaulieu08}
J.P. {Beaulieu}, E.~{Kerins}, S.~{Mao}, et~al., arXiv:0808.0005

\bibitem{covone00}
G.~{Covone}, R.~{de Ritis}, M.~{Dominik}, A.A. {Marino}, \aap ~\textbf{357}, 816
  (2000)

\bibitem{baltz01}
E.A. {Baltz}, P.~{Gondolo}, \apj ~\textbf{559}, 41 (2001)

\bibitem{chung06}
S.J. {Chung}, D.~{Kim}, M.J. {Darnley}, et~al., \apj ~\textbf{650}, 432 (2006)

\bibitem{ingrosso09}
G.~{Ingrosso}, S.~{Calchi Novati}, F.~{de Paolis},  et~al., 
\mnras ~\textbf{399}, 219 (2009)

\bibitem{novae04}
M.J. {Darnley}, M.F. {Bode}, E.~{Kerins}, et~al., \mnras
  ~\textbf{353}, 571 (2004)

\bibitem{novae06}
M.J. {Darnley}, M.F. {Bode}, E.~{Kerins}, et~al., \mnras 
 ~\textbf{369}, 257 (2006)

\bibitem{ansari04}
R.~{Ansari}, M.~{Auri{\`e}re}, P.~{Baillon}, et~al., \aap 
~\textbf{421}, 509 (2004)

\bibitem{fliri06}
J.~{Fliri}, A.~{Riffeser}, S.~{Seitz}, R.~{Bender}, \aap ~\textbf{445}, 423  (2006)

\bibitem{joshi03}
Y.C. {Joshi}, A.K. {Pandey}, D.~{Narasimha}, R.~{Sagar},
  Y.~{Giraud-H{\'e}raud}, \aap ~\textbf{402}, 113 (2003)

\bibitem{joshi04}
Y.C. {Joshi}, A.K. {Pandey}, D.~{Narasimha}, Y.~{Giraud-H{\'e}raud},
  R.~{Sagar}, J.~{Kaplan}, \aap ~\textbf{415}, 471 (2004)

\bibitem{gould95b}
A.~{Gould}, \apj,~ \textbf{455}, 44 (1995)

\bibitem{baltz04}
E.A. {Baltz}, T.R. {Lauer}, D.R. {Zurek}, {et~al.}, 
\apj,~ \textbf{610}, 691 (2004)

\bibitem{totani03}
T.~{Totani}, \apj,~ \textbf{586}, 735 (2003)

\bibitem{tuntsov04}
A.V. {Tuntsov}, G.F. {Lewis}, R.A. {Ibata}, J.~{Kneib}, \mnras ~\textbf{353},
  853 (2004)

\bibitem{dejong08}
J.T.A. {de Jong}, K.H. {Kuijken}, P.~{H{\'e}raudeau}, \aap,~ \textbf{478}, 755
  (2008)

\bibitem{baillon92}
P.~{Baillon}, A.~{Bouquet}, Y.~{Giraud-H{\'e}raud}, J.~{Kaplan}, in
  \emph{Particle Astrophysics}, ed. by {G.~Fontaine \& J.~Tran Thanh van}
  (1993), pp. 529--+
%Particle Astrophysics, Proceedings of the 4th Rencontres de Blois, Chateau de Blois, France, June 15-20, 1992. Edited by G. Fontaine and J. Tran Thanh Van. Gif-sur-Yvette: Editions Frontieres, 1993., p.529

\bibitem{kaplan92}
Paul Baillon, Alain Bouquet, Yannick Giraud-H\'eraud, and Jean Kaplan,
%\newblock Searching brown dwarfs by the microlensing of unresolved stars.
\newblock In G\'erard Fontaine and Jean {Tr\^an Thanh V\^an}, editors,
\emph{Fourth ``Rencontres de Blois'': Particle Astrophysics}, 
page 528, Gif sur Yvette, 1992, Editions Fronti\`eres



\bibitem{bouquet94}
A.~{Bouquet}, J.~{Kaplan}, A.L. {Melchior}, Y.~{Giraud-H{\'e}raud},
  P.~{Baillon}, in \emph{The Dark Side of the Universe - Experimental Efforts
  and Theoretical Framework} (1994), pp. 61--68, arXiv:astro-ph/9312009

\bibitem{gillieron96}
D.~Gillieron, AGAPE, internal note, Coll{\`e}ge de France, Paris, (1996)

\bibitem{kaplan97}
J. Kaplan, \emph{Pixel Lensing}, in \emph{Topics on Gravitational Lensing}
\newblock Napoli Series In Physics and Astrophysics, Bibliopolis (1998)

\bibitem{ledu00}
Y.~{Le~Du}, Ph.D. thesis, Coll{\`e}ge de France, Paris, (2000) 

\bibitem{agape95a}
R.~{Ansari}, M.~{Auri{\`e}re}, P.~{Baillon}, {et~al.}, 
  Nuclear Physics B Proceedings Supplements \textbf{43}, 165 (1995).
%Trends in Astroparticle Physics, Stockholm, Sweden, 22-25 Sep 1994


\end{thebibliography}

%\newpage

\begin{appendix}

\section{Brief history of the AGAPE group's beginning\footnote{P.~Baillon, 
A.~Bouquet, Y.~Giraud-H\`eraud, and J.~Kaplan, 
private communication.}}

The AGAPE collaboration has been formed in the early 1990.
The first and original idea of using M31 as a target
for microlensing observation is due to P.~Baillon.
He developed this interest after a seminar
given by M.~Spiro in 1990 who was then moving,
together with J.~Rich,
the first steps to build the microlensing EROS collaboration.
On April 25, 1991, P. Baillon held a seminar at the
LPNHE\footnote{Ecole Polytechnique, Palaiseau, France.}
on the possibility to detect microlensing events in M31
using a photomultiplier then in use, Themistocle. 
Y. Giraud-H\'eraud was taken by this idea and
involved A.~Bouquet and J.~Kaplan.
Even if not yet working on microlensing, they all  were 
aware of the ongoing  EROS project. In particular
it was soon realised that CCDs were a more suitable
instrument to carry out this observational programme.
The conclusions of the first preliminar analyses
have been presented by P.~Baillon in June 1992 \cite{baillon92}
and by J.~Kaplan in July of the same year \cite{kaplan92}.
A first work  was then submitted for a refereed
publication in early November 1992 \cite{agape93}. 
(The authors were not aware of the parallel work
carried out by A.~Crotts \cite{crotts92}, which,
still in the pre-arXiv era, had been published
only a few days before).
The technical aspects for the analysis of the observational data
have been developed  by A-L. Melchior \cite{bouquet94}. 
A key ingredient of the superpixel photometry,
the ``seeing stabilisation'' technique,
has been first elaborated by  D.~Gillieron \cite{gillieron96},
presented in \cite{kaplan97}
and then discussed more thoroughly in the PhD thesis of Y.~Le~Du \cite{ledu00}.
Finally, the observational campaign was begun
in 1994 using the 2m TBL telescope at Pic du Midi,
to be continued for further 2 years. 
Three members of the EROS collaboration, 
R.~Ansari, C.~Coutures and M.~Moniez,  
participated in AGAPE, bringing their know how
both in observations and in the development of analysis algorithms.

Coming to the name's choice, AGAPE. 
This comes from a suggestion
of A.~Bouquet, with an underlying idea 
of a (friendly) opposition with EROS (a name chosen
in opposition to the other microlensing project,
MACHO, whose name, on the other hand, had been thought
to recall the other dark matter candidate, the WIMP). 
Indeed, according to the ancient greek mythology,
AGAPE/EROS is the couple of holy versus profane love.
The right acronym was then found
to be ``Andromeda Galaxy and Amplified Pixel Experiment''.
The name AGAPE 
has been first presented in Stockholm in 1994 par R.~Ansari \cite{agape95a}. 

\end{appendix}

\end{document}